\documentclass{aa}
\usepackage{hyperref}
\usepackage{natbib} 
\usepackage{booktabs,graphicx,blindtext}
\usepackage{listings}
\usepackage{color} 
\usepackage{algorithm}
\usepackage{algpseudocode}
\usepackage{bm}
\usepackage{graphicx}
\usepackage{nameref}

\renewcommand{\algorithmicrequire}{\textbf{Input:}}
\renewcommand{\algorithmicensure}{\textbf{Output:}}

\begin{document}

\renewcommand{\arraystretch}{1.2}

\titlerunning{Deep Transfer Learning for Classification of Variable Sources}
\title{Deep Transfer Learning for Classification of Variable Sources}

\authorrunning{D.-W. Kim et al.}
\author{Dae-Won Kim\inst{1,}\thanks{dwk@etri.re.kr}
\and
Doyeob Yeo\inst{1,}\thanks{yeody@etri.re.kr}\and
Coryn A.L. Bailer-Jones\inst{2}\and
Giyoung Lee\inst{1}
}
\institute{
ETRI}
\institute{
Electronics and Telecommunications Research Institute (ETRI), 218 Gajeong-ro, Daejeon, Yuseong-gu, South Korea
\and
Max-Planck Institute for Astronomy (MPIA), K\"{o}nigstuhl 17, D-69117 Heidelberg, Germany}

\abstract
{Ongoing or upcoming surveys such as Gaia, ZTF, or LSST will observe light-curves of billons or more astronomical sources. This presents new challenges for identifying interesting and important types of variability. Collecting a sufficient number of labelled data for training is difficult, however, especially in the early stages of a new survey. Here we develop a single-band light-curve classifier based on deep neural networks, and use transfer learning to address the training data paucity problem by conveying knowledge from one dataset to another. First we train a neural network on 16 variability features extracted from the light-curves of OGLE and EROS-2 variables. We then optimize this model using a small set (e.g. 5\%) of periodic variable light-curves from the ASAS dataset in order to transfer knowledge inferred from OGLE/EROS-2 to a new ASAS classifier. With this we achieve good classification results on ASAS, thereby showing that knowledge can be successfully transferred between datasets. We demonstrate similar transfer learning using Hipparcos and ASAS-SN data. We therefore find that it is not necessary to train a neural network from scratch for every new survey, but rather that transfer learning can be used even when only a small set of labelled data is available in the new survey.}



\maketitle

\section{Introduction}
\label{sec:introduction}

In recent years, deep learning has achieved outstanding success in various research areas and application domains. For instance, Convolutional Neural Networks (CNNs) \citep{Lecun2015Natur.521..436L} use convolutional layers to regularize and build space-invariant neural networks to classify visual data \citep{Alex2012, GoodfellowNIPS2014, Szegedy2015}. AlphaGo, which is one of the most astonishing achievements of deep learning and is based on reinforcement learning, defeated several world-class Go players \citep{Silver2016Natur.529..484S}. In astronomy deep learning has become popular and has been used in many studies such as star-galaxy classification \citep{Kim2017MNRAS.464.4463K}, asteroseismology classification of red giants \citep{Hon2017MNRAS.469.4578H}, photometric redshift estimation \citep{DIsanto2018A&A...609A.111D}, galaxy-galaxy strong lens detection \citep{Lanusse2018MNRAS.473.3895L}, exoplanet finding \citep{Pearson2018MNRAS.474..478P}, point source detection \citep{Vafaei2019MNRAS.484.2793V}, and fast-moving object identification \citep{Duev2019MNRAS.486.4158D}, to name just a few examples.

Training a deep learning classification model requires a huge amount of labeled data which are not always readily available. This can be addressed using ``transfer learning'', a method that preserves prior knowledge inferred from one problem having sufficient samples (the ``source'') and apply it to another but related problem (the ``target'') \citep{Hinton2015arXiv150302531H, Yim8100237, Yeo8451121}. Transfer learning starts with a machine learning model that has been trained on a lot of labeled samples from the source. It then optimizes parameters in the model using a small number of labeled samples from the target. The source and target samples can be different but need to be related such that extracted features to train the model are general and relevant in both samples. Period, amplitude, and variability indices could be such features when the problem is one of classifying light-curves of variable stars.

Transfer learning has been used to solve a variety of problems such as medical image classification \citep{Shin2016arXiv160203409S, Maqsood2019, Michaldoi:10.1002/mrm.27969}, recommender-system applications \citep{Weike10.5555/2900728.2900823, Guangneng2019arXiv190107199H}, bioinformatics applications \citep{Xu5706537, Petegrosso10.1093/bioinformatics/btw649}, transportation applications \citep{Di7934021, Wang8120162, Lu8759907}, and energy saving applications \citep{Li6747280, Zhao7041046}. Some studies in astronomy also used transfer learning  such as \citet{Ackermann2018MNRAS.479..415A, Domnguez2019MNRAS.484...93D, Lieu2019MNRAS.485.5831L, Tang2019MNRAS.488.3358T}. These studies used transfer learning to analyze image datasets of either galaxies or solar system objects, and confirmed that transfer learning is successful even when using a small set of data.

The advantages of transfer learning are as follows:

\begin{itemize}

\item Transfer learning uses accumulated knowledge from the source. Therefore it works with a small set of samples from the target. 

\item Transfer learning uses a pre-trained model, obviating the need to design a deep neural network from scratch for every survey, which is a challenging and time-consuming task. A pre-trained model furthermore contains better initialization parameters than randomly initialized parameters. Thus it is likely that transferring a pre-trained model shows faster convergence than training from scratch.

\item It is possible to either add or remove certain output classes from the pre-trained model during transfer-learning processes. This is particularly useful because the classes in the target dataset are mostly different from the classes of the source dataset.

\end{itemize}

Our previous work, UPSILoN: A{\bf U}tomated Classification for {\bf P}eriodic Variable {\bf S}tars using Mach{\bf I}ne Lear{\bf N}ing \citep{Kim2016AA...587A..18K}, we provided a software package that automatically classifies light-curve into one of seven periodic variable classes. Other groups have used UPSILoN to classify light-curves of periodic variable stars (e.g. \citealt{Jayasinghe2018MNRAS.477.3145J, Kains2019MNRAS.482.3058K, Hosenie2019MNRAS.488.4858H}).

In this paper, we introduce UPSILoN-T: {\bf UPSILoN} using {\bf T}ransfer Learning. UPSILoN-T is substantially different from UPSILoN in several respects:

\begin{itemize}

\item UPSILoN-T can be transferred to other surveys whereas the UPSILoN's Random Forest model cannot be.

\item UPSILoN-T is not only able to classify periodic or semi-periodic variables but also non-periodic variables such as QSOs or blue variables. 

\item UPSILoN-T uses MCC (Matthews Correlation Coefficient; \citealt{MATTHEWS1975442}) rather than $F_1$ as a performance metric. MCC is known to give a relatively robust performance measure for imbalanced samples \citep{Boughorbel2017PLoS}. We use the generalized MCC \citep{Chicco:2017qb} for multi-class classification.

\item UPSILoN-T uses deep learning whereas UPSLIoN used a Random Forest. The UPSILoN-T model has three hidden layers with 64, 128, and 256 neurons, respectively.  This is a much smaller model ($\sim$200 KB) than UPSILoN's Random Forest model ($\sim$3 GB).

\end{itemize}

As far as we know, no previous work has applied transfer learning based on deep learning to time-variability features extracted from light-curves. 
In addition to developing the method, we provide a python library so that readers can easily adapt our method to their datasets (see \nameref{sec:appendix}). The library classifies a single-band light-curve into one of nine variable classes. If multiple optical bands exist, the library can be applied to each of them independently. Note that one of the goals of our method, as with its predecessor UPSILoN, is not to be dependent on colours and so to be independent of the availability or number of bands in the survey.


We give an introduction of a neural network in Section \ref{sec:deep_learning}. In section \ref{sec:training_set} we introduce the training set, and in section \ref{sec:variability_features} we explain 16 time-variability features that we use to train a classification model. From section \ref{sec:model_training} to \ref{sec:feature_importance}, we explain training processes and feature importance of the trained model. Section \ref{sec:transfer_learning_application_to_other_dataset} demonstrates the application of transfer learning to the other light-curve datasets: ASAS, Hipparcos, and ASAS-SN. Section \ref{sec:resampled} gives the classification performance of the transferred model as a function of light-curve lengths and sampling. Section \ref{sec:summary} gives a summary.

\section{Artificial Neural Network}
\label{sec:deep_learning}

An artificial neural network (ANN) consists of multiple layers comprising an input layer, hidden layer(s), and an output layer. ANN with at least two hidden layers are usually called a deep neural network (DNN). Each layer contains one or more nodes (neurons) that are connected to the nodes in the next layer. Each connection has a particular weight that can be interpreted as how much impact that node has on the connected node in the next layer. 
Data are propagated through the network and the value on each node is computed using an activation function \citep{agatonovic2000basic}. The Heaviside function, hyperbolic tangent function, and rectified linear unit (ReLU) \citep{nair2010rectified} are typical examples of nonlinear activation functions that are often used between intermediate layers. The softmax function is widely used in the output layer when solving a classification problem \citep{nair2010rectified}. Due to the nonlinearity, the function space that can be expressed using ANN is diverse.

\begin{figure}
\begin{center}
       \includegraphics[width=0.35\textwidth]{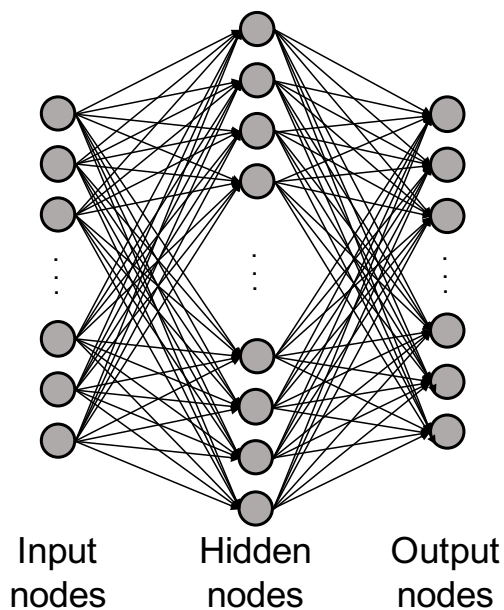}
\end{center}
    \caption{DNN architecture with one hidden layer.}
    \label{fig:dnn_simplest}
\end{figure}

An example of an ANN architecture is shown in Fig. \ref{fig:dnn_simplest}. We define input nodes, hidden nodes and output nodes as $\bm{x}$, $\bm{h}$ and $\bm{y}$, respectively, that are vectors. Let the weights and biases between $\bm{x}$ and $\bm{h}$, and weights between $\bm{h}$ and $\bm{y}$ be $W_1, \bm{b}_1$ and $W_2, \bm{b}_2$, respectively. Then, $W_1, W_2$ are matrices and $\bm{b}_1, \bm{b}_2$ are vectors. We can then express $\bm{x}$, $\bm{h}$, and $\bm{y}$ as follows:

\begin{equation}
\begin{array}{l}
\bm{h}=\sigma_1 \left(\bm{x}^T W_1 + \bm{b}_1 \right) \\
\bm{y}= \sigma_2 \left( \bm{h}^T W_2 + \bm{b}_2 \right)
\end{array} 
\label{eqn:dnn}
\end{equation}
where $\sigma_1(\cdot)$ and $\sigma_2(\cdot)$ denote the activation function between $\bm{x}$ and $\bm{h}$, and between $\bm{h}$ and $\bm{y}$, respectively. In classification problems, $\sigma_1$ and $\sigma_2$ are typically the ReLU function and the softmax function, respectively. The values of the output nodes give the probabilities that input (i.e. $\bm{x}$) belongs to each of the output classes. The goal of a classification problem is learning the weights so that the true classes receive the highest probabilities. Given an input $X = \left\{ \bm{x}_1, \bm{x}_2, \cdots , \bm{x}_m \right\}$, this learning process is done by minimizing a loss function such as cross entropy defined as follows:

\begin{equation}
\begin{array}{rcl}
 L(X;\bm{W}) & \equiv & \displaystyle -\frac{1}{m} \sum_{i=1}^{m} \log P(\bm{l}_i | \bm{y_i})\\
		     & = & \displaystyle -\frac{1}{m} \sum_{i=1}^{m} \bm{l}_i \cdot \log(\bm{y_i})
\end{array}
\label{eqn:crossentropy}
\end{equation}
where $\bm{l}_i$ the true label for each $\bm{x}_i$ and a one-hot encoded vector, which is a vector representation of categorical data such as class labels. That is, $\bm{l}_k = [l_{k,1}, l_{k,2}, \cdots , l_{k,c}]$ and $l_{k,j} \in \{ 0, 1\}$ for $k=\{1, 2, \cdots, m\}$ and $j=\{1, 2, \cdots, c\}$. $c$ is the number of classes. $\bm{W} = \left\{ W_1, \bm{b}_1, W_2, \bm{b}_2 \right\}$ and $\cdot$ denotes an inner product.
In the cases of imbalanced dataset classification, a weighted cross entropy is often used, which is defined as follows:

\begin{equation}
\begin{array}{ccl}
L(X;\bm{W}) & \equiv & \displaystyle -\frac{1}{m} \sum_{i=1}^{m} \bm{\alpha} \cdot (\bm{l}_i \circ \log( \bm{y_i}))\\
& = & \displaystyle -\frac{1}{m} \sum_{i=1}^{m}
\begin{bmatrix}
  \alpha_1  \\
  \alpha_2  \\
  \vdots \\
  \alpha_c
\end{bmatrix}
\cdot
\begin{bmatrix}
  l_{i,1} \log(y_{i,1})  \\
  l_{i,2} \log(y_{i,2})  \\
  \vdots \\
  l_{i,c} \log(y_{i,c})  \\
\end{bmatrix}
\end{array}
\label{eqn:wcrossentropy}
\end{equation}

{\noindent}where $\bm{\alpha} = [\alpha_1, \alpha_2, \cdots, \alpha_c]$ and $\circ$ denotes an element-wise product. $\alpha_i$ is a weight for a given class which is a nonnegative real number, where $\sum_{i=1}^{c}{\alpha_i}=1$.  

Inferring the weights (i.e. learning) is done by minimizing a loss function such as that in Equation~\eqref{eqn:crossentropy} or \eqref{eqn:wcrossentropy} using a gradient descent method: 
\begin{equation}
\bm{W}_{\textrm{new}} \leftarrow \bm{W}_{\textrm{old}} - \lambda \nabla_{\bm{W}} L(X; \mathbf{W}_\textrm{old}) 
\label{eqn:gradientdescent}
\end{equation}
where $\lambda$ is a learning rate for the gradient descent optimization which is a positive constant. 
The minimization is often done by using a stochastic gradient descent optimization (SGD) \citep{Hinton201210.1109, Graves2013arXiv1303.5778G} or a modified stochastic gradient descent optimization (\citealt{Kingma2014arXiv1412.6980K}). In the case of SGD, the input data are split into several batches (i.e. mini-batch) and then SGD is applied to each mini-batch as the pseudo-code shown in Algorithm~\ref{alg:dnnlearning}. 

\begin{algorithm}
\caption{ANN learning procedure}
\label{alg:dnnlearning}
\algorithmicrequire { \\
Initialized weights of ANN to be learned: $\bm{W} = \left\{ W_1, \bm{b}_1, W_2, \bm{b}_2 \right\}$
}\\
\algorithmicensure { \\
Learned weights of ANN: $\bm{W}$
}
\begin{algorithmic}[1]
\Statex
\While{$\bm{W}$ does not converge}
\State {Choose a mini-batch $X_{\mathcal{B}}$ of size $n$ \newline\null\hfill $X_{\mathcal{B}} = \left\{ \bm{x}_{\mathcal{B}, 1}, \cdots , \bm{x}_{\mathcal{B}, n} \right\} \subset X$} 
\State {$\bm{W}_\textrm{new} \leftarrow \displaystyle \bm{W}_\textrm{old} - \lambda \nabla_{\bm{W}} L_{\textrm{CE}}(X_\mathcal{B}; \mathbf{W}_\textrm{old}) $ \newline
\null\hfill\Comment{Update $\bm{W}$ using SGD} }
\EndWhile
\end{algorithmic}
\end{algorithm}

\section{UPSILoN-T Classifier}

\subsection{Training Set}
\label{sec:training_set}

\renewcommand{\arraystretch}{1.2}
\begin{table*}
\small
\begin{center}
\caption{The number of sources per true class in the OGLE and EROS-2 training set\label{tab:training_set}}
\begin{tabular}{rrrr}
\hline\hline
Superclass (Acronym) & Subclass & Number & Note \\
\hline
QSO & & 165 & \\
$\delta$ Scuti (DSCT) & & 3209 & \\
Type II Cepheid (T2CEPH) & & 300 & \\
Blue Variable (BV) & & 735 & \\
Non-variable (NonVar) & & 8050 &  \\
RR Lyrae (RRL) & &  & \\
		& ab & 19\,921 & \\
		& c & 4974 & \\
		& d & 1077 & \\
		& e & 1327 & \\
Cepheid (CEPH) & & &  \\
		& F & 2981 & fundamental \\
		& 1O & 2043 & first overtone \\
		& Other & 443 & \\
Eclipsing Binary (EB) &  &  & \\
		& EC & 1398 & contact \\
		& ED & 19\,075 &  detached \\
		& ESD &  7339 & semi-detached \\
Long Period Variable (LPV) & & &  \\
		& Mira AGB C & 1361 &  carbon-rich \\
		& Mira AGB O & 783 &  oxygen-rich \\
		& OSARG AGB & 25\,284 &  \\
		& OSARG RGB & 29\,516 &  \\
		& SRV AGB C & 6062 & carbon-rich \\
		& SRV AGB O & 8780 & oxygen-rich \\
\hline
Total & & 144\,823 & \\
\end{tabular}
\end{center}
\end{table*}

The training set consists of single-band light-curves of variable sources from two independent surveys, OGLE \citep{Udalski1997AcA....47..319U} and EROS-2 \citep{Tisserand2007AA...469..387T}. We mix variable sources from the two surveys in order to build a rich training set. The training set comprises non-variables and seven classes of periodic variables: $\delta$ Scuti stars, RR Lyraes, Cepheids, Type II Cepheids, eclipsing binaries, and long period variables. We also use subclasses (e.g. Cepheid F, 1O, and Other) of each class if these exist. These seven classes of variable sources are from \citet{Kim2016AA...587A..18K} who originally compiled and cleaned the list of periodic variable stars from several sources including \citet{Soszynski2008AcA....58..163S, Soszynski2008AcA....58..293S, Soszynski2009AcA....59....1S, Soszynski2009AcA....59..239S, Poleski2010AcA....60....1P, Graczyk2011AcA....61..103G, Kim2014AA...566A..43K}. As described in these papers, light-curves are visually examined and cleaned in order to remove light-curves that do not show variability. In the present work we also add to the training set QSOs and blue variables from \citet{Kim2014AA...566A..43K}. These two types of variables generally show non-periodic and irregular variability.  Note that the training set is not exhaustive since it does not contain all types of variable sources in the real sky. Thus a model trained on the training set will classify a light-curve that does not belong to the training classes into a {\em{most similar}} training class based on  variability.

The observation duration is about eight years for the OGLE light-curves, and about seven years for the EROS-2 light-curves. We use OGLE I-band light-curves and EROS-2 blue-band, $B_E$, light-curves because these generally have more data points and better photometric accuracy than OGLE V-band and EROS-2 red-band, $R_E$, respectively. The average number of data points in the light-curves is about 500.
Table \ref{tab:training_set} shows the number of objects per class in the training set. There is a total of 21 variable classes.

\subsection{Variability Features}
\label{sec:variability_features}

We extract 16 variability features from each light-curve in the training set. These features, listed in Table \ref{tab:variability_features}, are taken from \citet{Neumann1941, shapiro1965analysis, Lomb1976ApSS, Ellaway1978, Grison1994AA, Stetson1996PASP, Long2012PASP..124..280L, Kim2014AA...566A..43K, Kim2016AA...587A..18K}. These features have proven to be useful for classifying both periodic and non-periodic variables (e.g. QSOs) in our previous works \citep{Kim2011ApJ...735...68K, Kim2012ApJ...747..107K, Kim2014AA...566A..43K, Kim2016AA...587A..18K}. They are generic and can be extracted from any single-band light-curves. Four of the features ($\psi^{\eta}$,  $\psi^{CS}$, $m_{p90}$, and $m_{p10}$) are based on a phase-folded light-curve. We excluded survey-dependent features, such as colours, because we want to transfer accumulated knowledge from one survey to another. 

\renewcommand{\arraystretch}{1.2}
\begin{table*}
\small
\begin{center}
\caption{16 Variability features \label{tab:variability_features}}
\begin{tabular}{rlc}
\hline\hline
Feature & Description & Reference \\
\hline
Period & Period derived by the Lomb-Scargle algorithm & \citet{Lomb1976ApSS} \\
$\gamma_1$  & Skewness & \\
$\gamma_2$ & Kurtosis & \\
$\psi^{\eta}$ & $\eta$ \citep{Neumann1941} of a phase-folded light-curve & \citet{Kim2014AA...566A..43K} \\
$\psi^{CS}$ & Cumulative sum \citep{Ellaway1978} index of a phase-folded light-curve & \citet{Kim2014AA...566A..43K} \\
$Q_{3-1}$ & 3$^{rd}$ quartile (75\%) - 1$^{st}$ quartile (25\%) & \citet{Kim2014AA...566A..43K} \\
$A$ & Ratio of magnitudes brighter or fainter than the average magnitude & \citet{Kim2016AA...587A..18K} \\ 
$H_1$ & Amplitude derived using the Fourier decomposition & \citet{Grison1994AA}  \\
$m_{p90}$ & 90\% percentile of slopes of a phase-folded light-curve & \citet{Long2012PASP..124..280L} \\
$m_{p10}$ & 10\% percentile of slopes of a phase-folded light-curve & \citet{Long2012PASP..124..280L} \\
$R_{21}$ & 2$^{nd}$ to 1$^{st}$ amplitude ratio derived using the Fourier decomposition & \citet{Grison1994AA} \\
$R_{31}$ & 3$^{rd}$ to 1$^{st}$ amplitude ratio derived using the Fourier decomposition & \citet{Grison1994AA} \\
$\phi_{21}$ & Difference between 2$^{nd}$ and 1$^{st}$ phase  derived using the Fourier decomposition & \citet{Grison1994AA} \\
$\phi_{31}$ & Difference between 3$^{rd}$ and 1$^{st}$ phase derived using the Fourier decomposition & \citet{Grison1994AA} \\
$K$ & Stetson $K$ index & \citet{Stetson1996PASP} \\
$W$ & Shapiro--Wilk normality test statistics & \citet{shapiro1965analysis} \\
\hline
\end{tabular}
\end{center}
\end{table*}

{We take logarithm base 10 of nine of the features (period, $\gamma_1$, $\gamma_2$, $\psi^{\eta}$, $\psi^{CS}$, $Q_{3-1}$, $A$, $H_1$, and $m_{p90}$) because we found empirically that a model trained using these $\log_{10}$ applied features improved the performance over using the features directly. We scale each of the nine feature set as follows:

\begin{enumerate}

\item Derive the minimum value of a feature set.

\item Subtract the minimum value minus one from each feature in the feature set.


\item Take logarithm base 10 of the features.

\end{enumerate}

Performance did not improve when taking the logarithm of the other features, so they were used directly. Finally, we normalize each 16 feature set by calculating the standard score.

\subsection{Model Training}
\label{sec:model_training}

\subsubsection{Classification Performance Metric}
\label{sec:imbalanced_set}

In order to measure the classification performance of a trained model and then select the one that gives the best performance, we use the MCC (Matthews Correlation Coefficient; \citealt{MATTHEWS1975442}) metric rather than the mode traditional $F_1$ or accuracy. For two-class problems, the MCC, accuracy, and $F_1$ are defined as follows:

\begin{equation}
\text{\footnotesize{MCC}} =  \frac{\text{\footnotesize{\textit{TP $\times$ TN - FP $\times$ FN}}}}
{\sqrt{\text{\footnotesize{(\textit{TP + FP})(\textit{TP + FN})(\textit{TN + FP})(\textit{TN + FN})}}}}
\end{equation}

\begin{equation}
\text{Accuracy} = \frac{TP + TN}{TP + TN + FP + FN}
\end{equation}

\begin{equation}
F_1 = 2 \times \frac{\text{precision} \times \text{recall}}{\text{precision} + \text{recall}}
\end{equation}

{\noindent}where $TP, \, TN, \, FP, $ and $TN$ is the number of true positives, true negatives, false positives, and false negatives, respectively. Precision and recall in the equation of $F_1$ are defined as follows:

\begin{equation}
\text{precision} = \frac{TP} {TP + FP}\,\,,  \,\,\,\, \text{recall} = \frac{TP} {TP + FN}
\end{equation}

The MCC is a more informative measure for imbalanced datasets than the other two measures because the MCC utilizes all four categories of a confusion matrix \citep{Powers2011evaluation, Boughorbel2017PLoS}. $F_1$ does not take account of true negative and thus is less informative for imbalanced datasets. Accuracy is inappropriate for imbalanced datasets because high accuracy does not necessarily indicate high classification quality, due to the ``Accuracy Paradox'' \citep{FERNANDES2010338, Valverde-Albacete:2014hc}. The MCC ranges from -1.0 to 1.0, where 1.0 indicates that the predictions match exactly with the true labels, 0.0 indicates that the predictions are random, and -1.0 means that the predictions are completely opposite to the true labels. It is claimed that the MCC is the most informative value to measure classification quality using a confusion matrix \citep{Chicco:2017qb}. In the case of multi-class problems, the MCC is generalized as follows \citep{Gorodkin:2004:CTK}:

\begin{equation}
\begin{aligned}
\text{MCC} = \left[ \sum_k \sum_l \sum_m C_{kk} C_{lm} - C_{kl} C_{mk} \right] \\
	\times \left[ \sum_k \left( \sum_l C_{kl} \right) \left( \sum_{k^\prime | k^\prime \neq k} \sum_{l^\prime} C_{k^\prime l^\prime} \right) \right]^{- \frac{1}{2}} \\
	\times \left[ \sum_k \left( \sum_l C_{lk} \right) \left( \sum_{k^\prime | k^\prime \neq k} \sum_{l^\prime} C_{l^\prime k^\prime} \right) \right]^{- \frac{1}{2}}
\end{aligned}
\end{equation}

For convenience, we map MCC to the range [0, 1] as suggested in \citealt{Chicco:2020it}:

\begin{equation}
\begin{aligned}
M_{c} = \frac{1 + \text{MCC}} {2} 
\end{aligned}
\end{equation}

{\noindent}Hereinafter, we use $M_c$ to report classification quality of neural networks.

\subsubsection{Training DNN Models}
\label{sec:training_dnn_model}

\begin{figure*}
\begin{center}
       \includegraphics[width=0.8\textwidth]{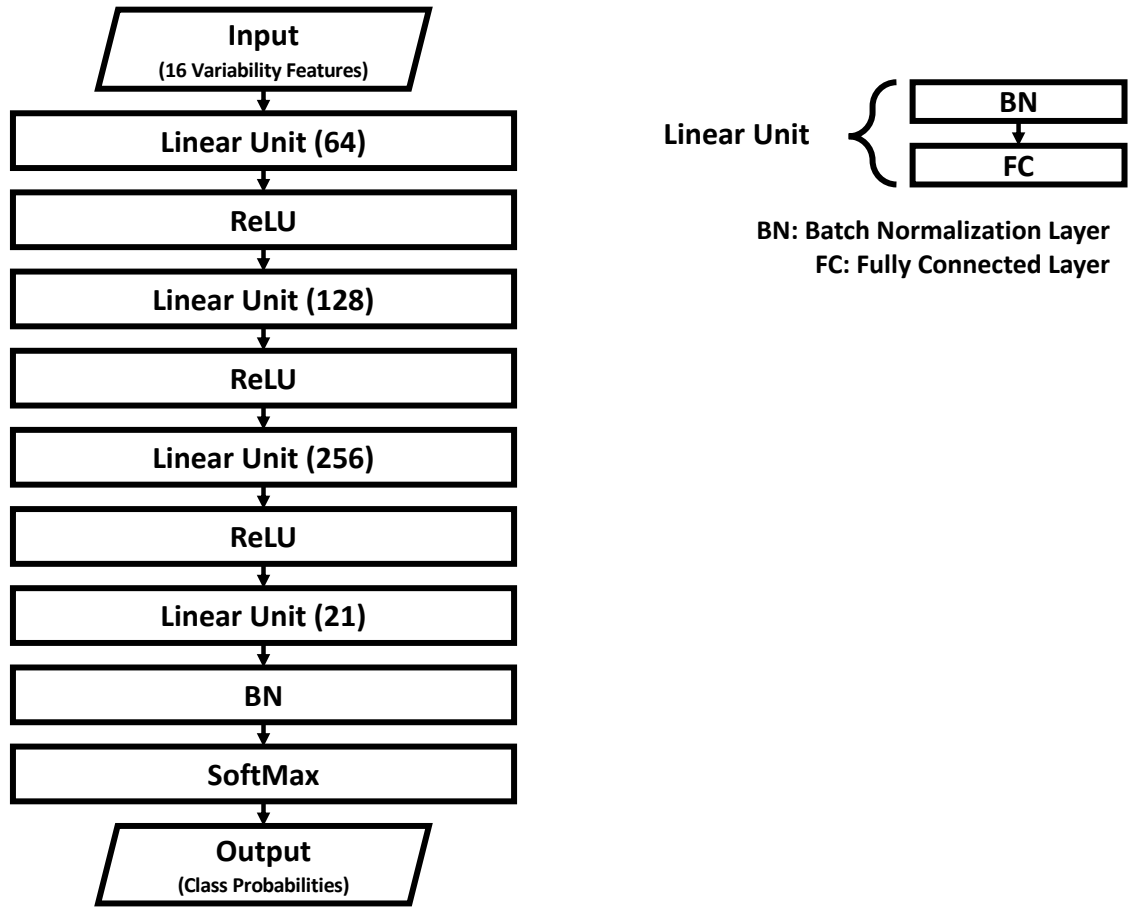}
\end{center}
    \caption{The neural network architecture that we adopted in this work. The input is a vector of 16 variability features. Each linear unit contains one batch normalization layer and one fully-connected layer. We use ReLU as an activation function between layers. The last linear unit has 21 outputs, which is the number of variable classes in our training set. We use the softmax function before the output layer in order to scale the outputs to lie between 0 and 1, to represent probabilities for the 21 output classes.}
    \label{fig:dnn_architecture}
\end{figure*}

\begin{figure}
\begin{center}
       \includegraphics[width=0.5\textwidth]{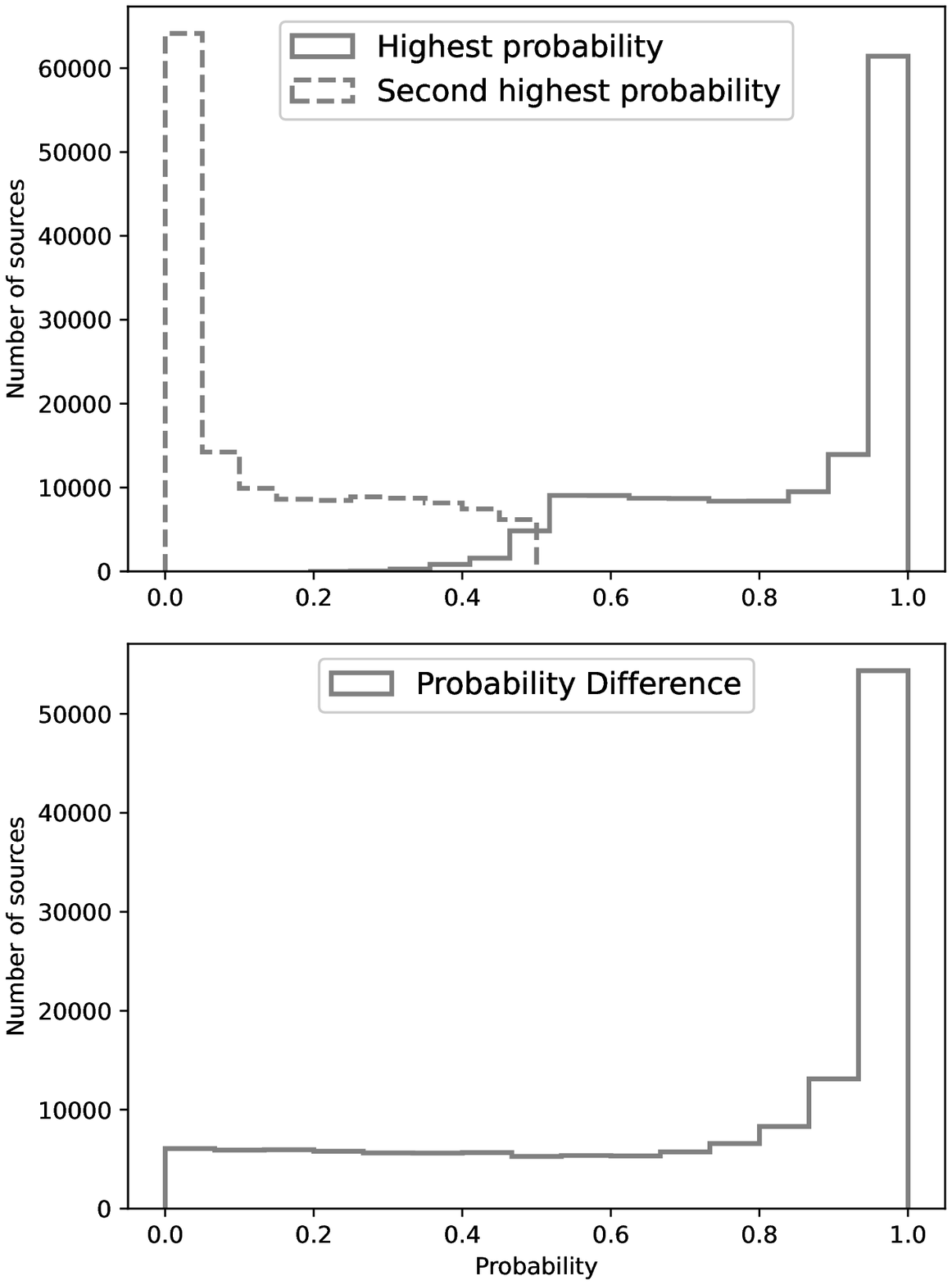}
\end{center}
    \caption{Class probability distribution of the light-curves in the training set. The upper panel shows the highest and the second highest probabilities. The bottom panel shows the differences between the two.}
    \label{fig:prob_hist}
\end{figure}


In order to find the best architecture for classifying the OGLE and EROS-2 training set, we trained many DNN models with different combinations of the number of hidden layers, the number of neurons in each hidden layer, loss functions, and activation functions between layers. The model training processes are as follows:

\begin{enumerate}

\item \label{item:start} Randomly split the dataset shown in Table \ref{tab:training_set} into 80\% training and 20\% validation set while preserving class ratio (i.e. stratified sampling).

\item \label{item:one_epoch} Train a DNN model from scratch using the training set with the batch size of 1024 and learning rate = 0.1. This counts as one training epoch. $M_c$ is then calculated using the validation set.

\item \label{item:end} Run the training process in step \ref{item:one_epoch} for 500 iterations (i.e. 500 training epochs). If $M_c$ is not improved for three consecutive epochs, decrease the learning rate by 10. If the learning rate becomes lower than $10^{-10}$, increase it back to 0.1, according to a technique called the  cyclical learning rate (CLR; \citealt{Smith2015arXiv150601186S}).

\item \label{item:best} Get the highest $M_c$ value (i.e. the best $M_c$ value) during the 500 epochs.

\item \label{item:repeat} Repeat steps 1 to 4 30 times and calculate the mean of the best $M_c$ values and the standard error of this mean (SEM), which is defined as:

\begin{equation}
\label{eqn:SEM}
\text{SEM} = \frac{\sigma}{\sqrt{n}}
\end{equation}

{\noindent}where $\sigma$ is standard deviation of the n $M_c$ values from the $n$=30 cycles.

\end{enumerate}

The loss function that we chose is cross entropy explained in Section \ref{sec:deep_learning}. 
While training DNN models, we use the SGD (stochastic gradient descent; \citealt{Hinton201210.1109}, \citealt{Graves2013arXiv1303.5778G}) optimizer because we empirically found SGD gives better classification performance than other optimizers (e.g. ADAM: \citealt{Kingma2014arXiv1412.6980K}; ADADELTA: \citealt{Zeiler2012arXiv1212.5701Z}; and  ADAGRAD: \citealt{JohnJMLR:v12:duchi11a}). We also use a batch normalization method that converts the inputs to layers into a normal distribution in order to mitigate ``internal covariate shift'' \citep{Ioffe2015arXiv150203167I}. \citet{Ioffe2015arXiv150203167I} showed that the method makes training faster and more stable.

After training many DNN models, we finally chose the architecture that have the highest validation-set $M_c$, which was 0.9043. This is shown in Fig. \ref{fig:dnn_architecture}. Note that we use $M_c$ to choose the best model but not as a loss function to optimize neural networks. We will refer to this best model as the ``U model'', an abbreviation of ``UPSILoN-T model''. The number of trainable parameters in the U model is 48,799. Given that the number of training samples is 144,823 and there are 16 features per sample, we presume that there exist enough features to train the parameters. Furthermore, the parameters in a network are not, and need not be, independent. Therefore it is not theoretically necessary to have more training features than the number of trainable parameters.

Some examples of the DNN architectures that we trained are shown in Table \ref{tab:DNN_model_performance} where the U model is shown in bold text.  As the table shows, increasing the number of layers and the number of neurons in each hidden layer increases classification performance. On the other hand, using too many neurons degrades classification performance as can be seen in the  bottom three rows due to overfitting of the training data. The $M_c$ differences between the models are statistically significant at the several sigma level. Whether the differences are {\em{practically}} significant or not is, however,  another issue. In order to confirm that the differences are practically significant, we would need to apply every trained DNN model to many different time-series databases having labels for light-curves and then check their classification performances. We have not done this, as we just chose the best model, but this does not preclude that a simpler model is in fact almost as good in practice.

We classify each light-curve according to the highest probability of all the classes. The top panel in Fig. \ref{fig:prob_hist} shows the highest and the second highest probabilities for each source in the training sample. As the panel shows, the majority of the highest probabilities are larger than 0.8 while the majority of the second highest probabilities are lower than 0.2. In the bottom panel, we show the histogram of differences between the two. As expected, the differences are generally larger than 0.5. This means that the classifier is reasonably confident about its class assignments. The U model returns both the predicted class and the probabilities of all classes, so users can set a threshold on a probability if they desire.

\renewcommand{\arraystretch}{1.2}
\begin{table*}
\small
\begin{center}
\caption{Classification performance of the neural network models \label{tab:DNN_model_performance}}
\begin{tabular}{rcccc}
\hline\hline
The Number of Hidden Layers  & Best $M_c$ & Avg. of $M_c$ & SEM$^2$ & Model Size \\
and Neurons in Each Layer$^1$ & & & & \\
\hline
32 & 0.8878 & 0.8848 & 3.9$\times 10^{-4}$ & 9 KB \\
64 & 0.8926 & 0.8891 & 3.1$\times 10^{-4}$ & 14 KB \\
128 & 0.8940 & 0.8925 & 1.9$\times 10^{-4}$ & 24 KB \\
256 & 0.8957 & 0.8929 & 2.1$\times 10^{-4}$ & 44 KB \\
32 $\rightarrow$ 64 & 0.9005 & 0.8970 & 2.2$\times 10^{-4}$ & 21 KB \\
64 $\rightarrow$ 128 & 0.9017 & 0.8999 & 1.9$\times 10^{-4}$ & 55 KB \\
128 $\rightarrow$ 256 & 0.9027 & 0.9002 & 2.4$\times 10^{-4}$ & 171 KB \\
256 $\rightarrow$ 128 & 0.8996 & 0.8934 & 1.6$\times 10^{-4}$ & 169 KB \\
32 $\rightarrow$ 64 $\rightarrow$ 128 & 0.9027 & 0.9004 & 1.9$\times 10^{-4}$ & 63 KB \\
\textbf{64 $\rightarrow$ 128 $\rightarrow$ 256}  & \textbf{0.9043} & \textbf{0.9015} & $\boldsymbol {2.2 \times 10^{-4}}$ & \textbf{202 KB} \\
128 $\rightarrow$ 256 $\rightarrow$ 512 & 0.9031 & 0.9008 & 2.2$\times 10^{-4}$ & 726 KB \\
256 $\rightarrow$ 512 $\rightarrow$ 1024 & 0.9018 & 0.9002 & 1.5$\times 10^{-4}$ & 3 MB \\
512 $\rightarrow$ 1024 $\rightarrow$ 2048 & 0.9032 & 0.9001 & 2.4$\times 10^{-4}$ & 11 MB \\
\hline
\end{tabular}
\end{center}
$^1$ The integer values indicate the number of neurons in each hidden layer. For instance ``$32 \rightarrow 64$'' means that there are two hidden layers, which have 32 and 64 neurons.
\\$^2$ $n$ in Equation \eqref{eqn:SEM} is 30.
\end{table*}

\subsubsection{Classification Quality of the U model}

\begin{figure*}
\begin{center}
       \includegraphics[width=1.0\textwidth]{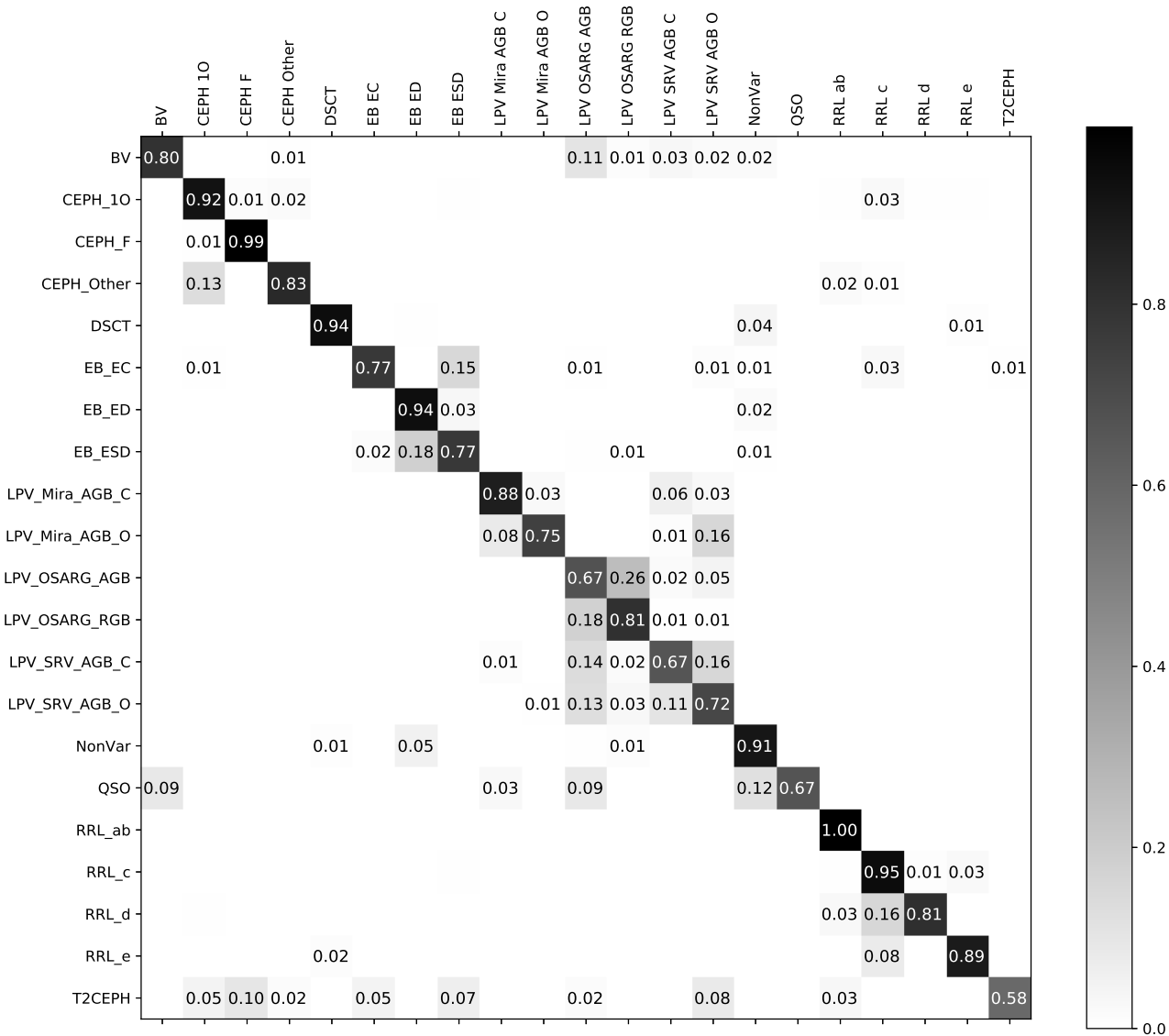}
\end{center}
    \caption{The confusion matrix of the U model. Labels on the vertical axis are the true classes and labels on the horizontal axis are the predicted classes.  Each row is divided by the number of true objects per variable class. Thus the values on each diagonal are recall. We show the value if it is larger than or equal to 0.01.}
    \label{fig:UT_confusion_matrix}
\end{figure*}

\begin{figure*}
\begin{center}
       \includegraphics[width=1.0\textwidth]{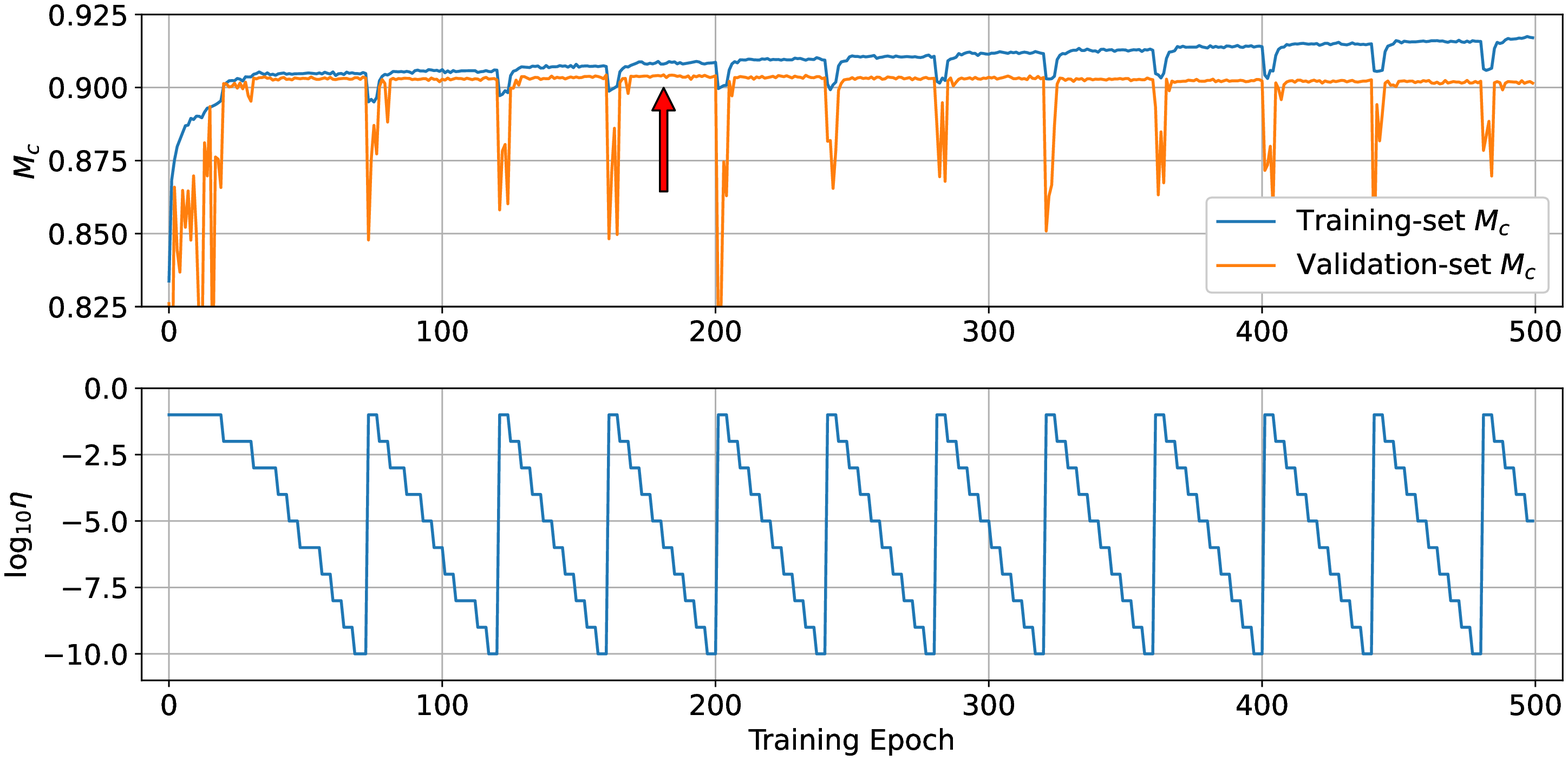}
\end{center}
    \caption{The top panel shows training-set $M_c$ (blue line) and validation-set $M_c$ (orange line) as a function of training epochs. The bottom panel shows changes of learning rate as a result of cyclical learning rate (CLR) application. The red arrow indicates the training epoch where the best validation-set $M_c$ is achieved.}
    \label{fig:training_info}
\end{figure*}

In Fig. \ref{fig:UT_confusion_matrix}, we show the U model's confusion matrix normalized by the number of true objects per class. The numbers on the leading diagonal represent recall values for each variable class. Most of the recall values are higher than $\sim$0.8 except for a few classes such as T2CEPH or QSO. The figure also shows confusion within subclasses of a certain variable class such as LPV or EB. This confusion is expected because variability patterns residing in their light-curves are likely to be similar.

In the top panel of Fig. \ref{fig:training_info}, we show $M_c$ and learning rate as a function of training epoch. The figure shows 500 epochs of the $n$=18 cycle (see step \ref{item:repeat} in Section \ref{sec:training_dnn_model}) where the highest validation-set $M_c$ (annotated with the red arrow) is achieved. As the figure shows, the training-set $M_c$ keeps increasing as a function of epoch. This indicates that the model eventually starts to overfit the training samples. Note that the U model is chosen based on the validation-set $M_c$, not based on the training-set $M_c$. The lower panel of Fig. \ref{fig:training_info} shows the change in the learning rate during the training processes. We see the learning rate cyclying decreasing and increasing, a consequence of our using CLR. $M_c$ value (the top panel) derived using the training sets and validation sets vary according to the changes of learning rate. The highest validation-set $M_c$ is obtained after several iterations of CLR. 

For comparison purposes, we trained Random Forest \citep{Breiman2001} models with the same dataset and variability features. We trained the models using 80\% of the data as a training set and then derived $M_c$ using the remaining 20\% data, just as we did for the DNN model training. We optimized the model hyper-parameters using the 80\% training set and using ten-fold cross-validation and brute-force grid search over a two-dimensional grid of $t$ and $m$, where $t$ is the number of trees and $m$ is the number of randomly selected features for each node in trees. The grid values of $t$ and $m$ are [100, 200, 300, 400, 500, 600, 700, 800, 900, 1000, 1100, 1200] and [4, 6, 8, 10, 12, 14], respectively.\footnote{The grid values are from our previous work,  \citet{Kim2016AA...587A..18K}.}. We repeated the training processes ten times. The highest $M_c$ after the training is 0.8996, which is lower than DNN's $M_c$. The average of $M_c$ and SEM\footnote{In this case, $n$ in the equation \ref{eqn:SEM} is 10.} are 0.8963 and 7.7$\times 10^{-4}$, respectively. Thus we concluded that the Random Forest model gives slightly worse performance than the U model. Nevertheless, this result does not mean that DNN is always superior than Random Forest for variable source classification using variability features. 

In the case of {\em{random}} light-curves that do not belong to any of the training classes, we found empirically that most of them are classified as non-variables as follows:

\begin{itemize}

\item We generated 1,000 light-curves consisting of purely random values, extracted 16 variability features from them, and predicted their classes using the U model. The U model predicted $\sim$800 of them as non-variables, $\sim$100 as long-period variables. The remaining $\sim$100 light-curves' predicted classes were dispersed through other variable classes.

\item We randomly shuffled data points in each light-curve of the ASAS variable stars (Table \ref{tab:n_class_asas} in Section \ref{sec:asas_application}), extracted 16 variability features, and used the U-model to predict their classes. As a result, $\sim$7,000 light-curves were classified as non-variables, and $\sim$3,000 light-curves were classified as EB EDs. The remaining $\sim$1,000 light-curves' predicted classes were dispersed through other classes.

\end{itemize}

\subsection{Application to ASAS light-curves of Periodic Variable Stars}
\label{sec:asas_application}

To validate the classification quality of the U model in general, we applied it to the ASAS light-curves of periodic variable stars that contain $\delta$ Scuti stars, RR Lyraes, Cepheids, eclipsing binaries, and long-period variables \citep{Pojmanski1997AcA....47..467P}. The duration of the light-curves is about nine years. The average number of data points is about 500. We collected the light-curves from \citet{Kim2016AA...587A..18K}. The number of samples per true variable class and the classification quality is shown in Table \ref{tab:n_class_asas}. We derived the $F_1$ measure of the same dataset as well, and it was 0.87. This is slightly higher than $F_1$=0.85 from \citet{Kim2016AA...587A..18K} that used the same training set and the 16 variability features to train a Random Forest. 
Nevertheless, this does not necessarily mean that the U model is always superior to the Random Forest. For instance, we applied the U model to light-curves from the MACHO dataset that \citet{Kim2016AA...587A..18K} also used. In this case, the $F_1$ measures for both the U model and  the Random Forest model are identical at 0.92.

\renewcommand{\arraystretch}{1.2}
\begin{table}
\small
\begin{center}
\caption{The number of objects per true class in the ASAS dataset \label{tab:n_class_asas}}
\begin{tabular}{cccc}
\hline\hline
Superclass & Subclass & Number & $M_c$\\
\hline
EB & & & \\
	& EC & 2550 & 0.91 \\
	& ED & 2167 & 0.96 \\
	& ESD & 823 & 0.86 \\
RRL & & & \\
	& ab & 1218 & 0.97 \\
	& c & 305 & 0.82 \\
CEPH & & & \\
	& F & 567 & 0.84 \\
	& 1O & 179 & 0.78 \\
DSCT & & 520 & 0.88 \\
LPV && 2450 & 0.97 \\
\hline
 Total   & & 10\,779 & 0.91 \\
\end{tabular}
\end{center}
\end{table}


\subsection{Feature Importance}
\label{sec:feature_importance}

We use the SHAP (\textbf{SH}apley \textbf{A}dditive ex\textbf{P}lanations; \citealt{LundbergNIPS2017_7062}) method to measure the feature importance in the U model. SHAP values represents how strongly each feature impacts on model's predictions. They are derived from the Shapley value \citep{Shapley1953qd} used in cooperative game theory. Conceptually, the Shapley value measures the degree of contribution of each player to a game where all the players cooperate. 
Although there are few SHAP applications in physics, \citet{Pham9006542, Amacker2020arXiv200404240A, Scillitoe2020arXiv200301968S} used SHAP to measure feature importances for their machine learning applications such as predicting solar flare activities, Higgs self-coupling measurements, and data-driven turbulence modelling.

Fig. \ref{fig:feature_importance} shows the SHAP feature importance for the U model. This shows the mean absolute value of SHAP over all samples for each feature. We see that the period is the most important feature, which has also been found in other studies (e.g. \citealt{Kim2014AA...566A..43K, Kim2016AA...587A..18K, Elorrieta2016AA...595A..82E}). All other features also contribute to the classification of variable classes although their contributions are lower than period's contribution.

We trained two independent DNN models with the same architecture as shown in Figure \ref{fig:dnn_architecture}. Both are trained with periods removed, the second with the period-related features ($\psi^{\eta}$,  $\psi^{CS}$, $m_{p90}$, and $m_{p10}$) also removed. As a result, the $M_c$ drops from 0.90 to 0.87 in the first case and to 0.82 in the second case. Given that the SEM values (Table \ref{tab:DNN_model_performance}) are quite small, these decreases are significant, again confirming that periods are critical for variable classification. Note, however, that periods of some training classes do not have a strict physical meaning as a period such as for quasars. In such cases, they are just used as a variability feature.

\begin{figure}
\begin{center}
       \includegraphics[width=0.53\textwidth]{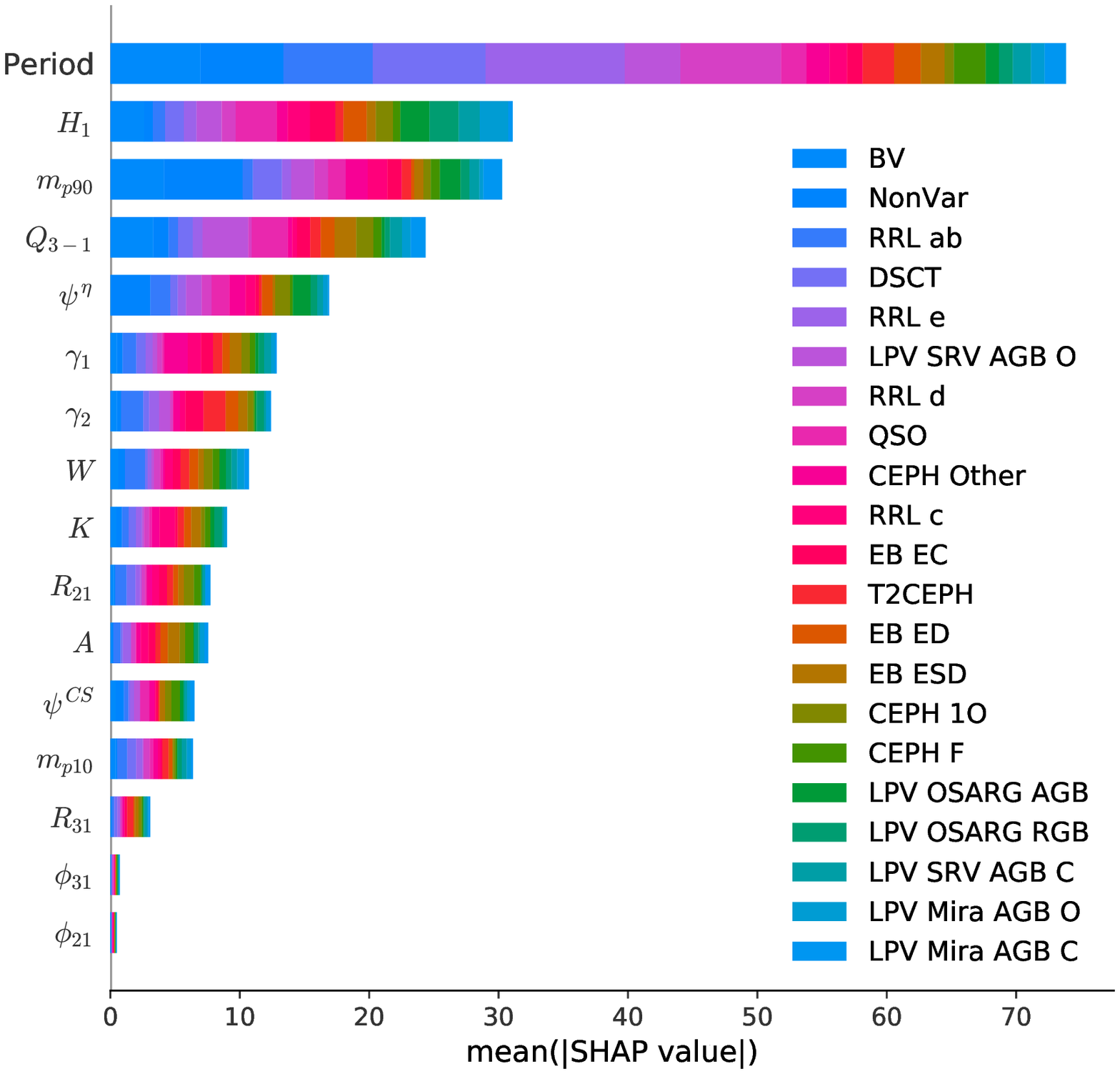}
\end{center}
    \caption{SHAP (Shapley additive explanations) feature importance \citet{LundbergNIPS2017_7062}. The larger the SHAP value, the more important a feature. The mean absolute SHAP value shows the accumulated feature importance for all variable classes depicted in different colors.}
    \label{fig:feature_importance}
\end{figure}

\section{Transferring the U model's Knowledge to Other Light-Curve Datasets of Variable Sources}
\label{sec:transfer_learning_application_to_other_dataset}

\subsection{Transfer Learning}
\label{sec:transfer_learning_process}

Transfer learning is a research field in machine learning that aims to preserve knowledge accumulated while solving one problem (source), then to utilize this to solve other but related problem (target) \citep{cowan1994advances, Pan:2010:STL:1850483.1850545}. 
Depending on the problem at hand, conditions of source and target could vary in several ways. In this work, which aims to transfer the U model to data from other surveys, we want to deal with the following issues:

\begin{enumerate}

\item Different class label space.
\\The number of variable classes (21 as shown in Table \ref{tab:training_set}) of the OGLE and EROS-2 training set is different from the number of variable classes of the ASAS dataset (9 as shown in Table \ref{tab:n_class_asas}). Moreover, not only the number of variable classes but also the variable types are different. For instance, there are no QSOs or blue variables in the ASAS dataset.

\item Different frequency of class members.
\\The number of contacted eclipsing binaries (EC) in the OGLE and EROS-2 training set  is 1398 (see Table \ref{tab:training_set}), which is less than 1\% of the dataset, but their number in the ASAS dataset is 2550 (see Table \ref{tab:n_class_asas}), which is $\sim$24\% of the dataset.

\item Different distribution of variability features.
\\Magnitude ranges, filter wavelengths, and noise characteristics are different per survey. This implies that the distribution of 16 variability features extracted from light-curves is not identical between source and target.


\end{enumerate}

The widely-used transfer-learning approach to deal with these issues is to transfer the weights from the pre-trained model. That is, we start with the same neural network model with the same weights (i.e. the U model). We then continue with the training in one or two different ways. The first is to optimize the weights of every layer, and the second is to optimize the weights of only the last fully-connected layer while freezing the other layers. In Fig. \ref{fig:dnn_architecture}, the last fully-connected layer is in the fourth linear unit (i.e. the last linear unit). In both approaches we adjust the number of output nodes at the last linear unit so that the number of output nodes is equal to the number of variable classes from the target dataset.\footnote{Technically, we replace the fully-connected layer in the fourth linear unit with a new fully-connected layer having the same number of output nodes with the number of target's variable classes.} By doing this, we can obtain appropriate prediction results with regard to the target dataset's classes.

In order to show the benefit of transfer learning, we define a base model (step \ref{item:base_model} in the following list) as the non-transferred and directly trained model, and compare it to the transferred models (step \ref{item:transfer_model}). We will also compare a model directly trained on a small dataset (scratch models, step \ref{item:scratch_model}) with the transferred models. The process for the transfer learning that we apply to each target dataset (i.e. ASAS, Hipparcos, and ASAS-SN) is as follows:

\begin{enumerate}

\item \label{item:transfer_exp_start} Randomly split the target dataset into 80\% and 20\% while preserving the class ratio (stratified sampling). We call the 80\% dataset $D_{0.8}$, and the 20\% dataset the test set.

\item \label{item:base_model} Train a DNN model from scratch with the same architecture shown in Fig. \ref{fig:dnn_architecture}. We use the entire $D_{0.8}$ to train the model. We call this the {\em{base model}}. The training process is the same as the ones from steps \ref{item:start} to \ref{item:end} in Section \ref{sec:training_dnn_model}. We then calculate $M_c$ of the base model using the test set. Note that this base model does not use any accumulated knowledge from the U model (i.e. weights in the U model). 

\item \label{item:scratch_model} Randomly choose a 3\% subset of $D_{0.8}$ and then use this subset to train a DNN model from scratch, which has the same architecture as the one shown in Fig. \ref{fig:dnn_architecture}. We use the same training process in Section \ref{sec:training_dnn_model}. We call this the {\em{scratch model}}. Also note that we do not transfer any of the U model's accumulated knowledge. After the training, we compute $M_c$ for the scratch model on the test set.

\item \label{item:transfer_model} Use the same 3\% subset of $D_{0.8}$ used for training the scratch model to transfer the U model. As mentioned above, we transfer the U model according two different approaches 1) optimizing weights of every layers, and 2) optimizing weights of only the last fully-connected layer while freezing weights of all other layers. We call the former a {\em{every-layer model}}, and the latter an {\em{last-layer model}}. Note that both approaches start from the initial weights of the U model rather than from random initialization. We again use the same training process in Section \ref{sec:training_dnn_model} to optimize the weights. We compute the $M_c$ of these models using the test set.

\item \label{item:transfer_repeat} Repeat steps \ref{item:scratch_model} to \ref{item:transfer_model} using 5\%, 10\%, 15\%, 20\%, 50\%, and 100\% of $D_{0.8}$. Theoretically, classification quality of the base model should be identical to the one of the scratch model trained using 100\% of $D_{0.8}$ because the base model is also trained using the same 100\% of $D_{0.8}$.

\item Repeat steps \ref{item:transfer_exp_start} to \ref{item:transfer_repeat} 30 times and use these to compute the mean and SEM value of $M_c$.

\end{enumerate}

In the following three sections, we show and analyze the results of using this transfer learning procedure on three different datasets.

\subsection{Application to the ASAS Dataset}
\label{sec:tr_app_asas_dataset}


\begin{figure}
\begin{center}
       \includegraphics[width=0.5\textwidth]{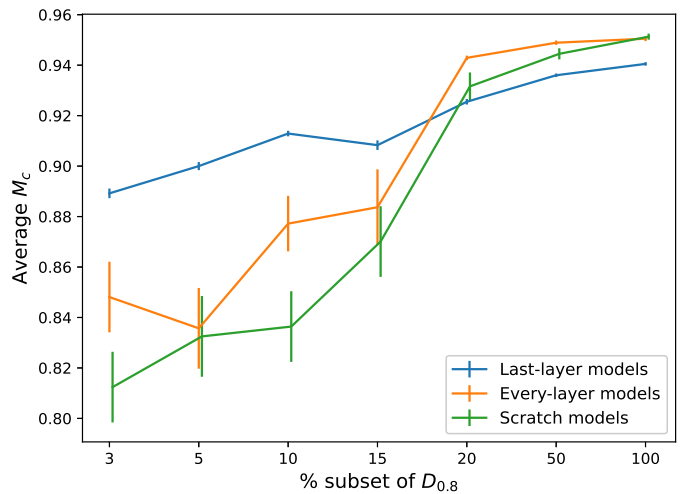}
\end{center}
    \caption{Average $M_c$ with SEM as error bars for the last-layer, every-layer, and scratch models. The every-layer models' $M_c$ are slightly higher than the scratch models' $M_c$. The last-layer models show higher $M_c$ for small training sets, and vice versa for large training sets. We add a small offset on the x-axis to the scratch models in order to prevent its error bars from lying on top of the other error bars.}
    \label{fig:asas_mc}
\end{figure}

The results of applying the transfer learning process to the ASAS dataset described in Section \ref{sec:asas_application} are given in Fig. \ref{fig:asas_mc}. The base model's average $M_c$ is 0.9501 and SEM is 1.8$\times 10^{-3}$.
The figure also shows the results for the other models (described in the previous section), which we now briefly discuss.

\begin{enumerate}

\item Scratch model.

The scratch models' $M_c$ increase as the size of a subset of $D_{0.8}$ increases. The one trained using 100\% of $D_{0.8}$ shows $M_c$ = 0.9513, which is very similar to the $M_c$ of the base model.

\item Every-layer model.

The every-layer models' $M_c$ are generally higher than the scratch models' $M_c$, even when their training set is small. This result shows the benefit of transfer learning, even when the target dataset is small. For larger samples (50\%-100\% subset), the results of the transferred models and scratch models are similar.

\item Last-layer model.

The last-layer models preserve more knowledge from the U model than the every-layer models do. For the same reason, the last-layer model could give worse performance because new knowledge from the target dataset cannot be transmitted to every layer. The results show that which effect dominates depends on the training sample size. 
For small training sets, the last-layer models produce a higher $M_c$, and vice versa for large training sets.
We also see that the SEM values of the last-layer models are smaller than other models, because optimizing the weights in only the last layer gives less freedom to update the weights than optimizing the weights in every layer.

\end{enumerate}

\begin{figure}
\begin{center}
       \includegraphics[width=0.5\textwidth]{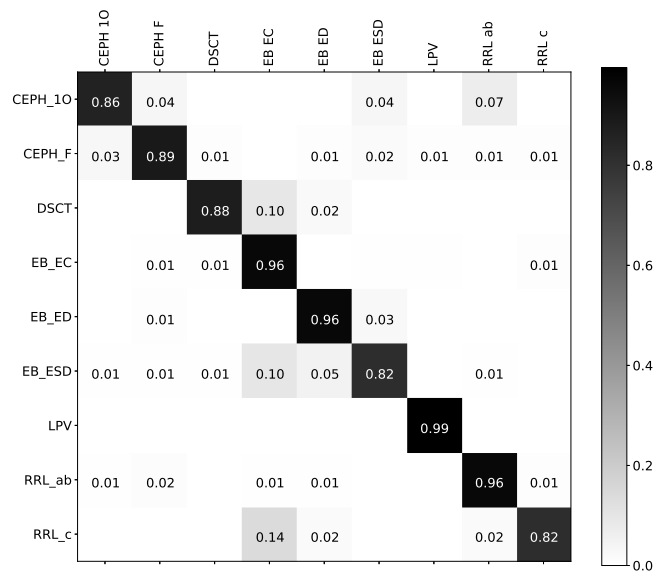}
\end{center}
    \caption{The confusion matrix of the ASAS transferred model. Labels are as in Table \ref{tab:n_class_asas}.}
    \label{fig:ASAS_confusion_matrix}
\end{figure}

Overall we see that transfer learning produces good results when the size of the target dataset is small, which is one of the advantages of using transfer learning. Transferring only the last layer is not a good choice when there is a relatively large number of target samples. In such cases, training from scratch (base model) or transferring every layer works better, although it requires more computation time. In addition, we show a confusion matrix of the every-layer model trained on the 100\% subset in Fig. \ref{fig:ASAS_confusion_matrix}. This shows good classification performance.

\subsection{Application to the Hipparcos Dataset}
\label{sec:tr_app_hip_dataset}

\renewcommand{\arraystretch}{1.2}
\begin{table*}
\small
\begin{center}
\caption{The number of objects per true class in the Hipparcos dataset \label{tab:n_class_hip}}
\begin{tabular}{lrrr}
\hline\hline
Type & Hipparcos Class & Number & UPSILoN-T class$^1$ \\
\hline
Eclipsing binary  & & \\
\,\,\,  detached & EA    &     222 &  EB ED \\
\,\,\,  semi-detached & EB    &     246  &  EB ESD \\
\,\,\,  contact & EW       &   99  & EB EC \\
RR Lyrae & RRAB      &  71 & RRL ab\\
& RRC   &      19  &  RRL c \\
$\delta$ Cepheid & & \\
\,\,\, fundamental mode & DCEP   &    170  & CEPH F \\
\,\,\, first overtone & DCEPS &      28  &  CEPH 1O \\
\,\,\, multi-mode & CEP(B)  &    11 & CEPH Other \\
$\delta$ Scuti & DSCT      &  45 & DSCT \\
\,\,\, low amplitude & DSCTC   &    80 & DSCT \\
Long-period variable & LPV    &    254 & LPV \\
Ellipsoidal & ELL     &    27 & \\
RS Canum Venaticorum & RS+BY     &  35  & \\
\,\,\, + BY Draconis & & & \\
Slowly pulsating B star & SPB      &   78  & \\
$\alpha$ Cygni & ACYG   &     17  & \\
$\alpha^2$ Canum Venaticorum & ACV      &   75  & \\
$\beta$ Cephei & BCEP      &  29  & \\
$\gamma$ Doradus & GDOR       & 25  & \\
B emmission line star & BE+GCAS &    12  & \\
\,\,\, + $\gamma$ Cassiopeiae & & & \\
W Virginis & & & \\
\,\,\, long-period (> 8 days) & CWA      &    9  & \\
\,\,\, short period  (< 8 days) & CWB    &      6  & \\
SX Arietis & SXARI    &    7  & \\
RV Tauri & RV        &   5  & \\
\hline
Total & & 1570 &  \\
\end{tabular}
\end{center}
$^1$ Corresponding classes in the training set.
\end{table*}

We now apply transfer learning to the Hipparcos dataset \citep{Dubath2011MNRAS.414.2602D} which has more classes but fewer samples than the ASAS dataset (see Table \ref{tab:n_class_hip}). The 1570 Hipparcos light-curves that we extracted from \citet{Kim2016AA...587A..18K} is substantially smaller than the 10\,779 light-curves in the ASAS dataset. The duration of the light-curves varies from one to three years, and the average number of data points is about 100.

Overall training process is the same as the one mentioned in Section \ref{sec:transfer_learning_process}. To ensure sufficient samples for training, however, we split the Hipparcos data into 50\%/50\% training/testing sets ($D_{0.5}$), rather than the 80\%/20\% that we used for ASAS. We use the entire set of $D_{0.5}$ to train scratch models and also to transfer the U model. We did not train separate base models since these are now the same as the scratch models.

The result of transfer learning is shown in Table \ref{tab:hip_transfer}. The scratch model's average $M_c$ is 0.7398 whereas the every-layer model's average $M_c$ is 0.7770. The last-layer model' achieves the highest average $M_c$ of all models tested, of 0.8230. Due to the small number of target samples, learning from scratch is significantly inferior to transfer learning in this case. Fig. \ref{fig:Hip_confusion_matrix} is a confusion matrix of the last-layer model, which shows relatively large confusion between the classes because of the insufficient target samples as well.


\begin{figure*}
\begin{center}
       \includegraphics[width=1.0\textwidth]{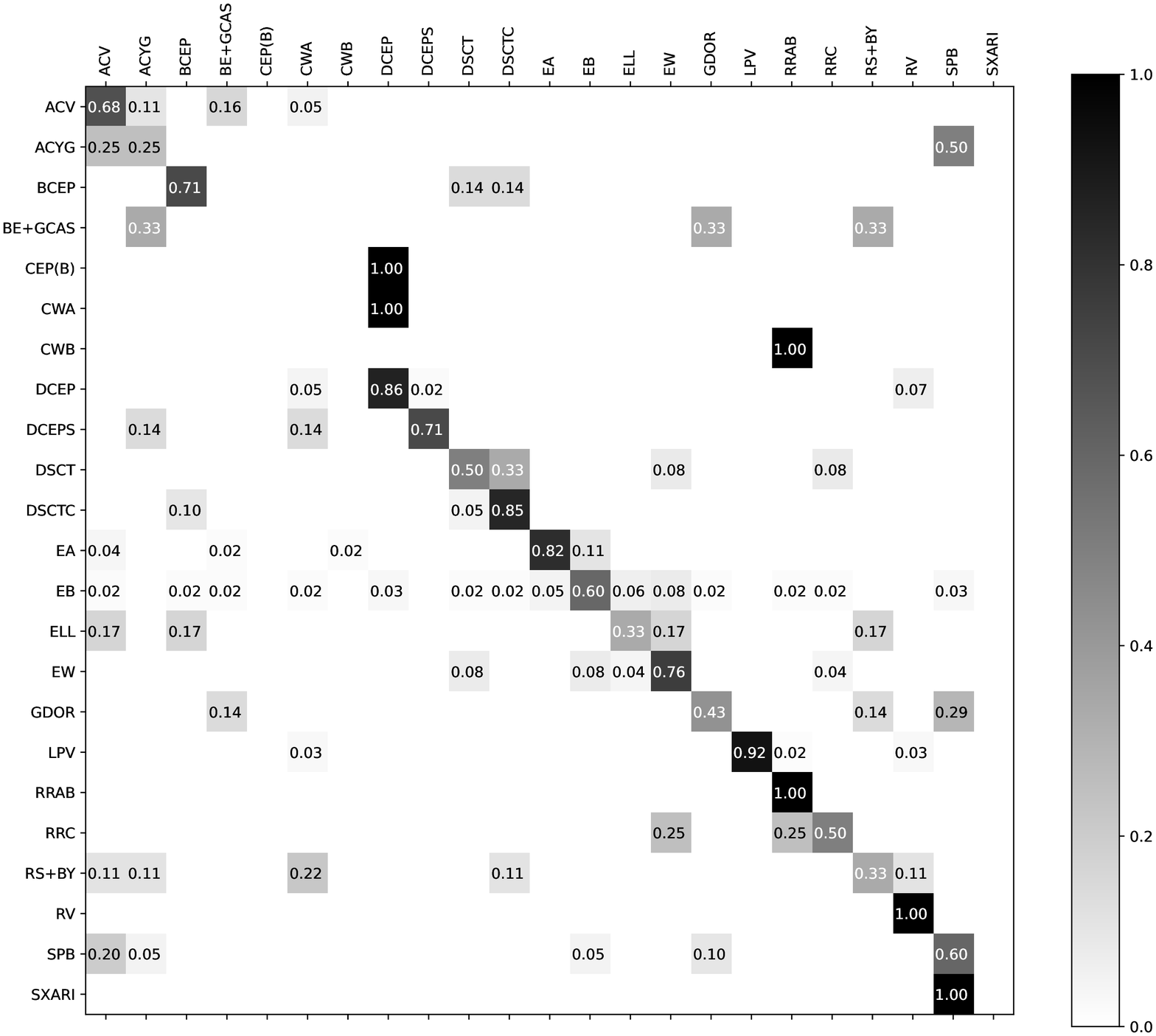}
\end{center}
    \caption{The confusion matrix of the Hipparcos transferred model. Labels are as in Table \ref{tab:n_class_hip}.}
    \label{fig:Hip_confusion_matrix}
\end{figure*}

\renewcommand{\arraystretch}{1.2}
\begin{table}
\small
\begin{center}
\caption{Transfer learning application using the Hipparcos dataset \label{tab:hip_transfer}}
\begin{tabular}{rcr}
\hline\hline
Model & Avg. of $M_c$ & SEM$^1$ \\
\hline
Scratch Model & & \\
 & 0.7398 & 1.3$\times 10^{-2}$ \\
\hline
Transfer Model & & \\
(every-layer model) & & \\
 & 0.7770 & 8.6$\times 10^{-3}$ \\
\hline
Transfer Model & & \\
(last-layer model) & & \\
 & 0.8230 & 1.8$\times 10^{-3}$ \\
\hline
\end{tabular}
\end{center}
$^1$ $n$ in Equation \eqref{eqn:SEM} is 30.
\end{table}

\subsection{Application to the ASAS-SN Dataset}
\label{sec:asas_sn}

\renewcommand{\arraystretch}{1.2}
\begin{table*}
\small
\begin{center}
\caption{The number of objects per true class in the ASAS-SN dataset \label{tab:n_class_asas_sn}}
\begin{tabular}{lrrr}
\hline\hline
Type & ASAS-SN Class & Number & UPSILoN-T class$^1$ \\
\hline
Cepheid & DCEP        &    743 & CEPH F \\ 
 $\delta$ Cephei-type variable & DCEPS        &    207 & CEPH 1O \\
 \,\,\, with a symmetrical light-curve & & & \\
$\delta$ Scuti & DSCT        &    1412 & DSCT \\
High amplitude $\delta$ Scuti & HADS        &    1901 & DSCT \\
RR Lyrae &         &     &  \\
\,\,\, with an asymmetric light-curve & RRAB        &    23\,891 & RRL ab \\
\,\,\, with a symmetric light-curve & RRC        &    6226 & RRL c \\
Double mode RR Lyrae & RRD        &    296 & RRL d \\
Eclipsing binary &         &     &  \\
\,\,\,  W Ursae Majoris type & EW        &    38\,284 & EB EC \\
\,\,\,  Detached Algol-type & EA        &    22\,483 & EB ED \\
\,\,\,  Beta Lyrae type & EB        &    10\,857 & EB ESD \\
W Virginis type variable & CWA    &    230  & T2CEPH \\
\,\,\, with period > 8 days & & & \\
Mira variable & M        &    7498 & LPV \\
Red irregular variable & L        &    53\,598 &   \\
Semi-regular variable & LSP        &    160 &  \\
Yellow semi-regular variable & SRD        &    135 &  \\
Young stellar object & YSO        &    189 &  \\
Suspected rotational variable & ROT        &    11\,395 &  \\
Suspected semi-regular variable & SR        &    108\,943 &  \\
Uncertain type of variable & VAR        &    250 &  \\
\hline
Total & & 288\,698 &  \\
\end{tabular}
\end{center}
$^1$ Corresponding classes in the training set.
\end{table*}

\renewcommand{\arraystretch}{1.2}
\begin{table}
\small
\begin{center}
\caption{Transfer learning application using ASAS-SN dataset \label{tab:asas_sn_transfer}}
\begin{tabular}{rcr}
\hline\hline
Model & Avg. of $M_c$ & SEM$^1$ \\
\hline
Scratch Model & & \\
 & 0.8744 & 7.7$\times 10^{-4}$ \\
\hline
Transfer Model & & \\
(every-layer model) & & \\
 & 0.8711 & 1.1$\times 10^{-3}$ \\
\hline
Transfer Model & & \\
(last-layer model) & & \\
 & 0.8380 & 2.6$\times 10^{-4}$ \\
\hline
\end{tabular}
\end{center}
$^1$ $n$ in Equation \eqref{eqn:SEM} is 30.
\end{table}

\begin{figure}
\begin{center}
       \includegraphics[width=0.5\textwidth]{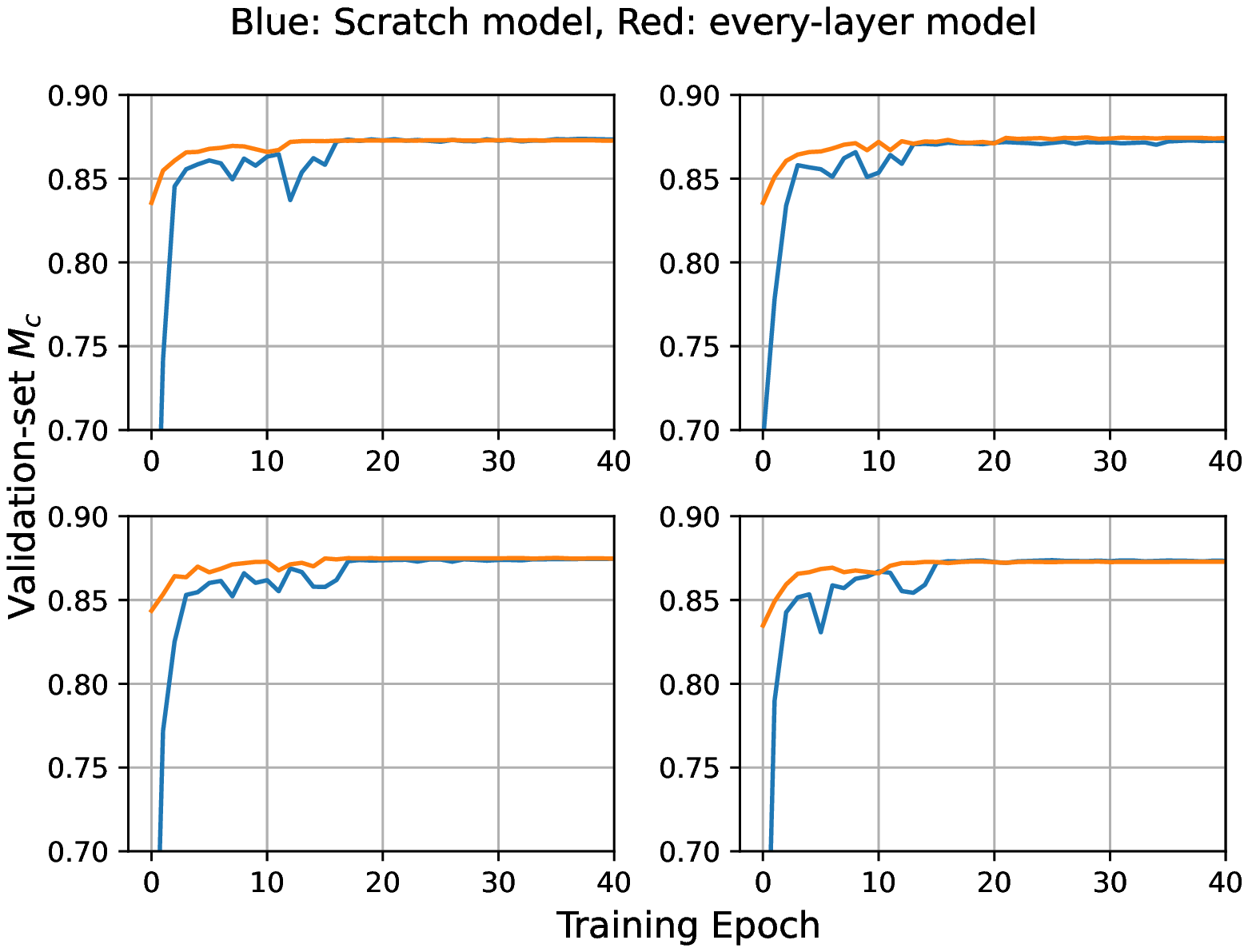}
\end{center}
    \caption{Examples of ASAS-SN validation-set $M_c$ of the scratch models (blue line) and the every-layer models (orange line). As the figure shows, transferring the UT model reaches the highest $M_c$ faster than training from scratch does.}
    \label{fig:asas_sn_mc}
\end{figure}

In this section, we apply transfer learning to the ASAS-SN database of variable stars \citep{Shappee2014ApJ...788...48S, Jayasinghe2019MNRAS.485..961J}. We obtained a catalog of the variable stars and their corresponding light-curves from the ASAS-SN website (\href{https://asas-sn.osu.edu/variables}{https://asas-sn.osu.edu/variables}). We selected only the light-curves with class probabilities higher than 95\%, and whose classification results are not uncertain. The average number of data points in the light-curves is about 200, and the duration is about four years. We extracted 16 time-variability features from each light-curve, excluding light-curves having fewer than 100 data points. In Table \ref{tab:n_class_asas_sn} we show the number of selected light-curves for each variable class and their corresponding classes\footnote{We manually assigned the corresponding classes.} in the training set (if any). The larger number of samples, 288\,698, is sufficient to train DNNs from scratch. Yet even in such cases transfer learning could be useful because it could achieve the highest $M_c$ faster than training from scratch does.

We train scratch, every-layer, and last-layer models using the entire set of $D_{0.8}$ according to the training process mentioned in Section \ref{sec:transfer_learning_process}. Because base models are now the same as the scratch models, we did not train them. The average of $M_c$ and SEM 
is shown in Table \ref{tab:asas_sn_transfer}. The average $M_c$ of the scratch model and the every-layer model are similar as expected, whereas that of the last-layer model is smaller. Note that we did not train models using a subset of the ASAS-SN dataset because its result would be similar to what we explained in Section \ref{sec:tr_app_asas_dataset} and \ref{sec:tr_app_hip_dataset}. Fig. \ref{fig:ASAS_SN_confusion_matrix} is a confusion matrix of the every-layer model. The classification quality is relatively good, but a lot of the SRD, VAR, YSO, LSP, and L are misclassified as SR. Most of these variables are either semi-regular or irregular variables, and thus their variability characteristics are expected to be similar to one another (i.e. no clear periodicities in their light-curves). Furthermore, the number of SRD, VAR, YSO, and LSP is relatively small, a few hundreds per class, whereas the number of SR is 108\,943, which could cause the bias toward the majority (SR).

Fig. \ref{fig:asas_sn_mc} shows four examples of validation-set $M_c$ at the first 40 training epochs from the $n$=30 cycles (see steps \ref{item:start} to \ref{item:repeat} in Section \ref{sec:training_dnn_model}). From the figure, we see that the every-layer models reach the highest $M_c$ faster than the scratch models do. We confirmed that the remaining 26 cases out of the $n$=30 cycles show similar behavior. However the difference is not significant, which implies that given enough training samples and computing resources, transfer learning is not really necessary because training from scratch gives as good classification performance as transfer learning does.


\begin{figure*}
\begin{center}
       \includegraphics[width=0.9\textwidth]{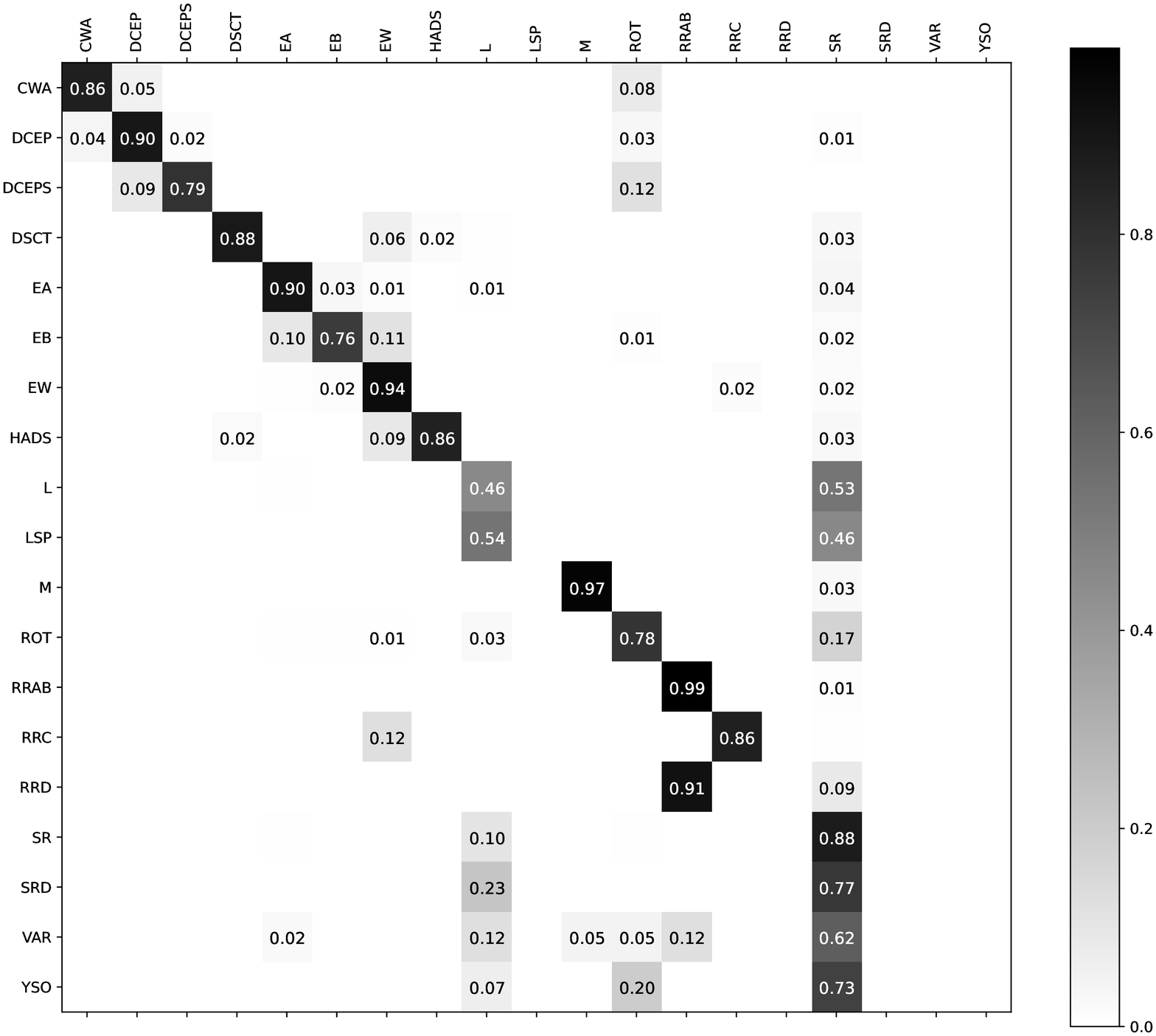}
\end{center}
    \caption{The confusion matrix of the ASAS SN transferred model. Labels are as in Table \ref{tab:n_class_asas_sn}.}
    \label{fig:ASAS_SN_confusion_matrix}
\end{figure*}

\section{Performance of a Transferred Model as a Function of Light-Curve Length and Sampling}
\label{sec:resampled}

\begin{figure*}
\begin{center}
\begin{minipage}[c]{9cm}
        \includegraphics[width=1.0\textwidth]{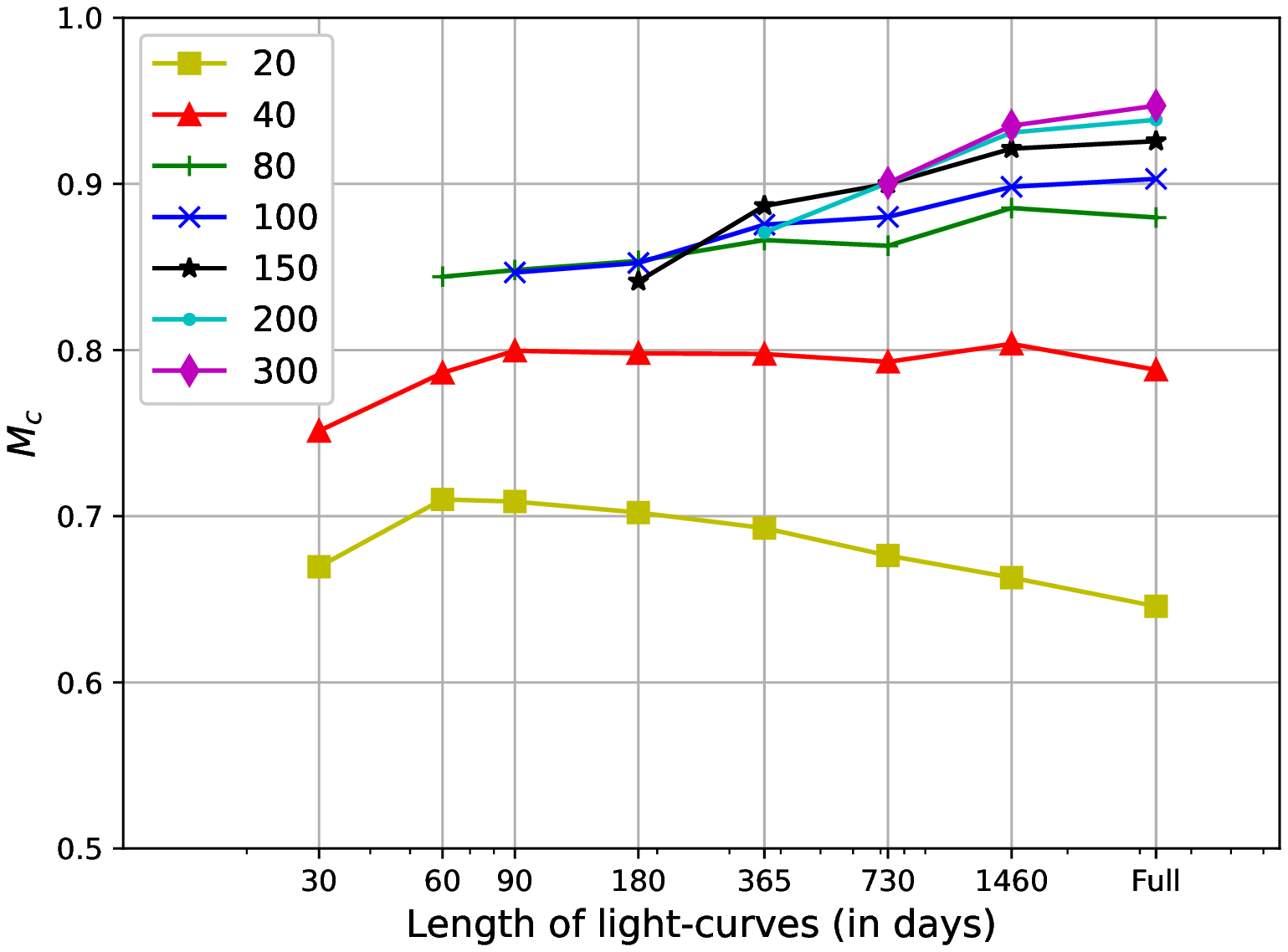}
\end{minipage}
\begin{minipage}[c]{9cm}
        \includegraphics[width=1.0\textwidth]{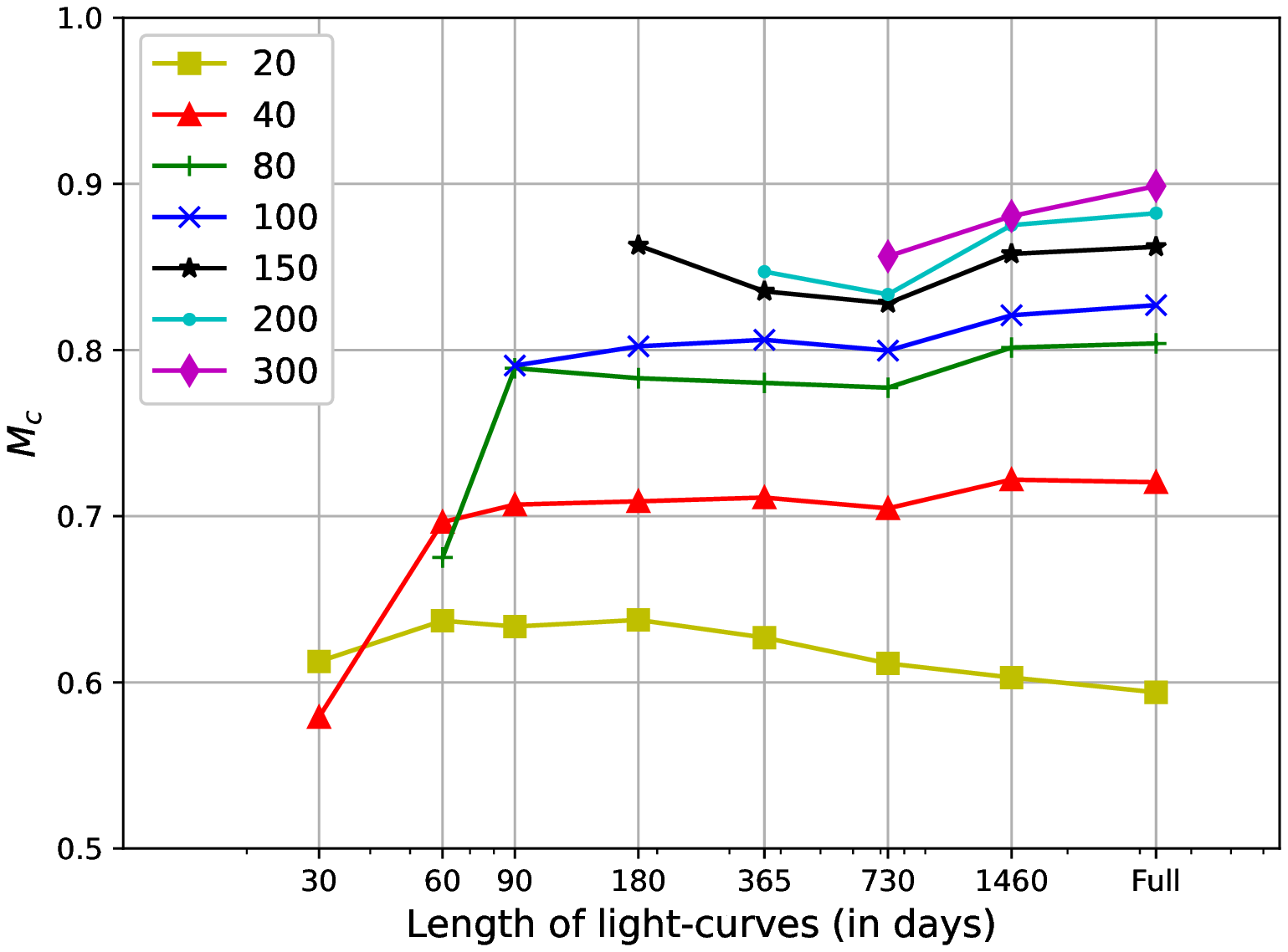}
\end{minipage}
\end{center}
    \caption
           {Classification quality, $M_c$, of the resampled ASAS light-curves as a function of duration, $d$ (horizontal axis), and the number of data points $n$ (different lines) in the light-curve. Left: $M_c$ of the transferred model. Right: $M_c$ of the U model.}
    \label{fig:mc_resampled}
\end{figure*}

In this section we examine the classification performance of the transferred model as a function of the observation durations, $d$, and the number of data points in a light-curve, $n$. For this experiment, we used the every-layer model transferred using the 20\% subset of ASAS $D_{0.8}$ (see Section \ref{sec:tr_app_asas_dataset}). $M_c$ of the model is 0.94.

We constructed a set of light-curves by resampling the ASAS light-curves shown in Table \ref{tab:n_class_asas} using $d$ = 30, 60, 90, 180, 365, 730, and 1470 days, and $n$ = 20, 40, 80, 100, 150, 200, and 300. In order to resample a given light-curve, we first extract measurements observed between the starting epoch and $d$ days after the starting epoch. We then randomly select $n$ unique data points from the extracted measurements to make a resampled light-curve. The left panel of Fig. \ref{fig:mc_resampled} shows the classification quality of the transferred model applied to these resampled light-curves. As the panel shows, $M_c$ quickly reaches its maximum value once the number of data points reaches 300. Even with lower $n$ = 100 or 150, or lower $d$ = 365 or 730, $M_c$ is fairly high. When $n$ = 20, we see that $M_c$ decreases as $d$ increases. This is because $n$ = 20 is too small number of data points to construct a well-sampled light-curve.

From this experiment, we see that if the number of data points, $n$, is larger than or equal to 100 or 150, $M_c$ is not significantly lower than what we can achieve using more data points. For comparison, we show $M_c$ of the U model for the same dataset in the right panel of Fig. \ref{fig:mc_resampled}. $M_c$ of the U model is lower than the transferred model, as expected, but the overall patterns of $M_c$ are similar with those of the transferred model. For the Hipparcos and ASAS-SN datasets, we did not resample their light-curves to carry out the same experiment due to a lack of either $n$ or $d$. Nevertheless, if enough $n$ and $d$ is given, we expect to see similar variations in $M_c$ for other datasets considered, because those models are also derived from the U model using transfer learning.

\section{Summary}
\label{sec:summary}


We have introduced deep transfer learning for classifying light-curves of variable sources in order to diminish difficulties in collecting sufficient training samples with labels, designing an individual neural network for every survey, and applying inferred knowledge from one dataset to another. Our approach starts with training a deep neural network using 16 variability features that are relatively survey-independent. These features are extracted from a training set composed of the light-curves of OGLE and EROS-2 variable sources. We then apply transfer learning to optimize the weights in the trained network so that we can extend the original model to new datasets, here ASAS, Hipparcos, and ASAS-SN. From these applications we see that transfer learning successfully utilizes knowledge inferred from the source dataset (i.e. OGLE and EROS-2) to the target dataset even with a small number of the target samples. In particular, we see that:

\begin{itemize}

\item Transferring only the last layer of the network shows better classification quality than transferring every layer when there are few samples. For instance, the last-layer models' classification performance is 3\%-7\% higher than the scratch models when a small number of ASAS samples (i.e. 3\%-15\% of the entire ASAS dataset) are used (see Fig. \ref{fig:asas_mc}).

\item As shown in Fig. \ref{fig:asas_mc}, when there exists a sufficient number of samples (i.e. 20\%-100\% of the entire ASAS dataset), the every-layer models give better classification performance 1\%-2\% higher than the last-layer models.

\item Transfer learning successfully works even when label space are different between the source and the target. For instance, the Hipparcos dataset contains many types of variable sources different from the training samples (i.e. OGLE and EROS-2) as shown in Table  \ref{tab:n_class_hip}. Transfer learning successfully works as the every-layer and the last-layer models show $\sim$4\% and $\sim$9\% better classification performance, respectively, than the scratch models (see Table \ref{tab:hip_transfer}).


\end{itemize}



From these results of using transfer learning, we see that transfer learning is useful for classification of variable sources from  time-domain surveys that have a few training samples. This is a significant benefit since it is not always possible to build a robust training set with a sufficient number of samples. 
For instance, many ongoing and upcoming time-domain surveys such as Gaia \citep{Gaia2016A&A...595A...1G}, ZTF \citep{Bellm2019PASP..131a8002B}, SkyMapper \citep{Keller2007PASA...24....1K}, and Pan-STARRS \citep{Chambers2016arXiv161205560C} will produce a large number of light-curves of astronomical sources. Assembling a sufficiently large set of labelled light-curves for training for each survey, however, is a difficult and time-consuming task, especially in the early stage of the survey. Given that transfer learning works even when using a small set of data, transfer learning would prove useful for classifying light-curves and thus building an initial training set for such surveys.
Nevertheless, if enough training samples are readily available, transfer learning does not give much benefit but at least it does not do any worse as can be seen from the results of every-layer models (e.g. Fig. \ref{fig:asas_mc} and Table \ref{tab:asas_sn_transfer}).

We have not used any survey-specific features such as colours. It would nonetheless be interesting to test whether including such features improves classification performance. For instance, one could replace the lowest important feature $\phi_{21}$ (see Fig. \ref{fig:feature_importance}) with a colour. One could then train a DNN model from scratch with this and transfer the trained model to another survey that uses different colours.

We provide a python software package containing a pre-trained network (the U model) and functionality to transfer the U model to any time-domain surveys. The package also can train a DNN model from scratch and deal with imbalanced datasets. 
For details about the package, see \nameref{sec:appendix}.


\section*{Acknowledgement}

This work was supported by a National Research Council of Science \& Technology (NST) grant by the Korean government (MSIP) [No. CRC-15-05-ETRI].

\bibliographystyle{aa}
\bibliography{bib}{}

\section*{Appendix}
\label{sec:appendix}

\definecolor{mygreen}{rgb}{0,0.6,0}
\definecolor{mygray}{rgb}{0.5,0.5,0.5}
\definecolor{mymauve}{rgb}{0.58,0,0.82}

\lstset{ %
  backgroundcolor=\color{white},   
  basicstyle=\footnotesize,        
  breakatwhitespace=false,         
  breaklines=true,                 
  captionpos=b,                    
  commentstyle=\color{mygray},    
  deletekeywords={...},            
  escapeinside={\%*}{*)},          
  extendedchars=true,              
  frame=single,                    
  keepspaces=true,                 
  keywordstyle=\color{blue}\bf,       
  language=Python,                 
  otherkeywords={*},            
  numbers=none,                    
  numbersep=8pt,                   
  numberstyle=\tiny\color{mygreen}, 
  rulecolor=\color{black},         
  showspaces=false,                
  showstringspaces=false,          
  showtabs=false,                  
  stepnumber=2,                    
  stringstyle=\color{mymauve},     
  tabsize=1,                       
  title=\lstname                   
}

\subsection*{How to Predict Variable Classes Using UPSILoN-T}

The UPSILoN-T python software package is available from \href{https://etrioss.kr/ksb/upsilon-t}{https://etrioss.kr/ksb/upsilon-t}\footnote{Permission is required to access the repository. Please send an email to one of the authors of this paper, either D.-W. Kim or D. Yeo.} The following pseudocode shows how to use the package to extract variability features from a set of light-curves and then to predict their variable classes.
        
\begin{lstlisting}[language=Python, frame=single, basicstyle=\tiny]
from upsilont import UPSILoNT
from upsilont.features import VariabilityFeatures

# Extract features from a set of light-curves. 
feature_list = []
for light_curve in set_of_light_curves:

    # Read a light-curve.
    date = np.array([:])
    mag = np.array([:])
    err = np.array([:])
         
    # Extract features.
    var_features = VariabilityFeatures(date, mag, err).get_features()
    
    # Append to the list.
    feature_list.append(var_features)

# Convert to Pandas DataFrame.
features = pd.DataFrame(feature_list)
    
# Classify.
ut = UPSILoNT()
label, prob = ut.predict(features, return_prob=True)
\end{lstlisting}

{\noindent}{\texttt{label}} and {\texttt{prob}} is a list of predicted classes and class probabilities, respectively. 

\subsection*{How to Transfer UPSILoN-T's Knowledge to Another Dataset}

The UPSILoN-T package can transfer the U model as follows:

\begin{lstlisting}[language=Python, frame=single, basicstyle=\tiny]
# Get features and labels.
features = ...
labels = ...

# Transfer UPSILoN-T.
ut = UPSILoNT()
ut.transfer(features, labels, "/path/to/transferred/model/")
\end{lstlisting}

{\noindent}\texttt{features} is a list of features extracted from a set of light-curves, and \texttt{labels} is a list of corresponding labels. The package writes the transferred model and other model-related parameters to a specified location (i.e. ``/path/to/transferred/model/'' in the above pseudocode). The transferred model can predict variable classes as follows:

\begin{lstlisting}[language=Python, frame=single, basicstyle=\tiny]
# Get features.
features = ...

# Predict.
ut = UPSILoNT()
ut.load("/path/to/transferred/model/")

label, prob = ut.predict(features, return_prob=True)
\end{lstlisting}


\subsection*{How to Deal with an Imbalanced Dataset}

As shown in Table \ref{tab:training_set}, training set is often imbalanced. For instance, the number of OSARG RGBs is about 179 times more than the number of QSOs (i.e. 29\,516 / 165 $\approx$ 179). Due to such class imbalance, training of a network could be biased towards the more dominant classes, even though this dominance is just a consequence of the relative frequency of available training samples that we do not want to learn.

There are several approaches to deal with imbalanced datasets such as weighting a loss function according to the sample frequencies, synthesizing artificial samples (e.g. SMOTE introduced at \citealt{Chawla.JAIR.2002}, or ADASYN introduced at \citealt{He2008ADASYNAS}), over- or under-sampling methods that repeat or remove samples in order to balance the sample frequency per class, or Bayesian methods \citep{Bailer-Jones2019MNRAS.490.5615B}. \citet{Bailer-Jones2019MNRAS.490.5615B} introduces a method to accommodate class imbalance in probabilistic multi-class classifiers. The UPSILoN-T library provides two of these approaches when training or transferring a model:

\begin{itemize}

\item Weighting a loss function

In Equation \ref{eqn:wcrossentropy}, we weight a loss function using {\bm{$\alpha$}}, which is proportional to the number of samples per class. In brief, we give higher weight to the minority class, and vice versa. {\bm{$\alpha$}} for each class is defined as follows:

\begin{equation}
\alpha_i = \frac{f_i} {\sum_{i=1}^{c} f_i} \,\, , \,\text{where} \,\, f_i = \frac{1}{N_i} \,\, ,
\end{equation}

{\noindent}where $N_i$ is the number of samples per class and $c$ is the number of variable classes.

\item Over-sampling technique

Conceptually, an over- or under-sampling method builds an artificial training set by balancing the number of each class in the original training set. Among them, the over-sampling method resamples the original training set by randomly duplicating samples from the minority class. For instance, if there are 20 samples of class A and 40 samples of class B, over-sampling randomly selects twice as many samples from class A as class B. 
The disadvantage of this approach to addressing imbalance is that the training time is increased. Another disadvantage is that a trained model could overfit the minority class since it just duplicates samples.

\end{itemize}


The following pseudocode shows how to use these two methods.

\begin{lstlisting}[language=Python, frame=single, basicstyle=\tiny]
ut = UPSILoNT()

# Over-sampling.
ut.train(features, labels, balanced_sampling=True)

# Weighting a loss function.
ut.train(features, labels, weight_class=True)

# Do both.
ut.train(features, labels, balanced_sampling=True, weight_class=True)
\end{lstlisting}


\end{document}